\documentclass[%
 aps,
prd,
twocolumn,showpacs,superscriptaddress,nofootinbib,floatfix]{revtex4-2}

\usepackage{graphicx}
\usepackage{dcolumn}
\usepackage{bm}

\usepackage{amsmath}   
\usepackage{graphicx}  
\usepackage{xspace}
\usepackage{units}
\usepackage{comment}
\usepackage{xcolor}

\usepackage{natbib}
\usepackage{hyperref}
\newcommand{\minerva}{MINER$\nu$A}

\newcommand{\gf}{\textsc{Geant4}\xspace}

\newcommand{\sizecheck}{0} 
\newcommand{\PRDsupp}{1}   
\ifnum\PRDsupp=0
  
\else
  
\fi

\begin{document}
\title{Constraint of the MINERvA Medium Energy Neutrino Flux using Neutrino-Electron Elastic Scattering} 

\newcommand{\Rutgers}{Rutgers, The State University of New Jersey, Piscataway, New Jersey 08854, USA}
\newcommand{\Hampton}{Hampton University, Dept. of Physics, Hampton, VA 23668, USA}
\newcommand{\Dortmund}{Institute of Physics, Dortmund University, 44221, Germany }
\newcommand{\Otterbein}{Department of Physics, Otterbein University, 1 South Grove Street, Westerville, OH, 43081 USA}
\newcommand{\JMU}{James Madison University, Harrisonburg, Virginia 22807, USA}
\newcommand{\Florida}{University of Florida, Department of Physics, Gainesville, FL 32611}
\newcommand{\UCIrvine}{Department of Physics and Astronomy, University of California, Irvine, Irvine, California 92697-4575, USA}
\newcommand{\CBPF}{Centro Brasileiro de Pesquisas F\'{i}sicas, Rua Dr. Xavier Sigaud 150, Urca, Rio de Janeiro, Rio de Janeiro, 22290-180, Brazil}
\newcommand{\PUCP}{Secci\'{o}n F\'{i}sica, Departamento de Ciencias, Pontificia Universidad Cat\'{o}lica del Per\'{u}, Apartado 1761, Lima, Per\'{u}}
\newcommand{\INRM}{Institute for Nuclear Research of the Russian Academy of Sciences, 117312 Moscow, Russia}
\newcommand{\Jlab}{Jefferson Lab, 12000 Jefferson Avenue, Newport News, VA 23606, USA}
\newcommand{\Pittsburgh}{Department of Physics and Astronomy, University of Pittsburgh, Pittsburgh, Pennsylvania 15260, USA}
\newcommand{\Guanajuato}{Campus Le\'{o}n y Campus Guanajuato, Universidad de Guanajuato, Lascurain de Retana No. 5, Colonia Centro, Guanajuato 36000, Guanajuato M\'{e}xico.}
\newcommand{\Athens}{Department of Physics, University of Athens, GR-15771 Athens, Greece}
\newcommand{\Tufts}{Physics Department, Tufts University, Medford, Massachusetts 02155, USA}
\newcommand{\WM}{Department of Physics, College of William \& Mary, Williamsburg, Virginia 23187, USA}
\newcommand{\FNAL}{Fermi National Accelerator Laboratory, Batavia, Illinois 60510, USA}
\newcommand{\Purdue}{Department of Chemistry and Physics, Purdue University Calumet, Hammond, Indiana 46323, USA}
\newcommand{\MCLA}{Massachusetts College of Liberal Arts, 375 Church Street, North Adams, MA 01247}
\newcommand{\UMD}{Department of Physics, University of Minnesota -- Duluth, Duluth, Minnesota 55812, USA}
\newcommand{\Northwestern}{Northwestern University, Evanston, Illinois 60208}
\newcommand{\UNI}{Universidad Nacional de Ingenier\'{i}a, Apartado 31139, Lima, Per\'{u}}
\newcommand{\Rochester}{University of Rochester, Rochester, New York 14627 USA}
\newcommand{\Austin}{Department of Physics, University of Texas, 1 University Station, Austin, Texas 78712, USA}
\newcommand{\USM}{Departamento de F\'{i}sica, Universidad T\'{e}cnica Federico Santa Mar\'{i}a, Avenida Espa\~{n}a 1680 Casilla 110-V, Valpara\'{i}so, Chile}
\newcommand{\Geneva}{University of Geneva, 1211 Geneva 4, Switzerland}
\newcommand{\Chicago}{Enrico Fermi Institute, University of Chicago, Chicago, IL 60637 USA}
\newcommand{\hired}{}
\newcommand{\OregonState}{Department of Physics, Oregon State University, Corvallis, Oregon 97331, USA}
\newcommand{\oxford}{Oxford University, Dept. of Physics, Oxford, United Kingdom}
\newcommand{\umiss}{University of Mississippi, Oxford, Mississippi 38677, USA}
\newcommand{\upenn}{Department of Physics and Astronomy, University of Pennsylvania, Philadelphia, PA 19104}
\newcommand{\AMU}{AMU Campus, Aligarh, Uttar Pradesh 202001, India}
\newcommand{\wroclaw}{University of Wroclaw, plac Uniwersytecki 1, 50-137 Wrocław, Poland}
\newcommand{\Mohali}{IISER, Mohali, Knowledge city, Sector 81, Manauli PO 140306}


\author{E.~Valencia}                      \affiliation{\WM}  \affiliation{\Guanajuato}
\author{D.Jena}                           \affiliation{\FNAL}
\author{Nuruzzaman}                       \affiliation{\Rutgers} 
   \affiliation{\USM}
\author{F.~Akbar}                         \affiliation{\AMU}
\author{L.~Aliaga}                        \affiliation{\WM}  \affiliation{\PUCP}
\author{D.A.~Andrade}                     \affiliation{\Guanajuato}
\author{M.~V.~Ascencio}                   \affiliation{\PUCP}
\author{A.~Bashyal}                       \affiliation{\OregonState}
\author{L.~Bellantoni}                    \affiliation{\FNAL}
\author{A.~Bercellie}                     \affiliation{\Rochester}
\author{A.~Bodek}                         \affiliation{\Rochester}
\author{J.~L.~Bonilla}                    \affiliation{\Guanajuato}
\author{A.~Bravar}                        \affiliation{\Geneva}
\author{H.~Budd}                          \affiliation{\Rochester}
\author{G.~Caceres}                       \affiliation{\CBPF}
\author{T.~Cai}                           \affiliation{\Rochester}
\author{M.F.~Carneiro}                    \affiliation{\OregonState}
\author{J.~Chaves}                        \affiliation{\upenn}
\author{D.~Coplowe}                       \affiliation{\oxford}
\author{H.~da~Motta}                      \affiliation{\CBPF}
\author{S.A.~Dytman}                      \affiliation{\Pittsburgh}
\author{G.A.~D\'{i}az~}                   \affiliation{\Rochester}  
\affiliation{\PUCP} 
\author{J.~Felix}                         \affiliation{\Guanajuato}
\author{L.~Fields}                        \affiliation{\FNAL}  
\author{A.~Filkins}                       \affiliation{\WM}
\author{R.~Fine}                          \affiliation{\Rochester}
\author{N.~Fiza}                          \affiliation{\Mohali}
\author{A.M.~Gago}                        \affiliation{\PUCP}
\author{R.Galindo}                        \affiliation{\USM}
\author{H.~Gallagher}                     \affiliation{\Tufts}
\author{A.~Ghosh}                         \affiliation{\USM}  \affiliation{\CBPF}
\author{T.~Golan}                         \affiliation{\wroclaw}  \affiliation{\Rochester}
\author{R.~Gran}                          \affiliation{\UMD}
\author{D.A.~Harris}                      \affiliation{\FNAL}
\author{S.~Henry}                         \affiliation{\Rochester}
\author{S.~Jena}                          \affiliation{\Mohali}
\author{J.~Kleykamp}                      \affiliation{\Rochester}
\author{M.~Kordosky}                      \affiliation{\WM}
\author{D.~Last}                          \affiliation{\upenn}
\author{T.~Le}                            \affiliation{\Tufts}  \affiliation{\Rutgers}
\author{X.-G.~Lu}                         \affiliation{\oxford}
\author{E.~Maher}                         \affiliation{\MCLA}
\author{S.~Manly}                         \affiliation{\Rochester}
\author{W.A.~Mann}                        \affiliation{\Tufts}
\author{C.~Mauger}                        \affiliation{\upenn}
\author{K.S.~McFarland}                   \affiliation{\Rochester}  \affiliation{\FNAL}
\author{A.M.~McGowan}                     \affiliation{\Rochester}
\author{B.~Messerly}                      \affiliation{\Pittsburgh}
\author{J.~Miller}                        \affiliation{\USM}
\author{J.G.~Morf\'{i}n}                  \affiliation{\FNAL}
\author{D.~Naples}                        \affiliation{\Pittsburgh}
\author{J.K.~Nelson}                      \affiliation{\WM}
\author{C.~Nguyen~}                       \affiliation{\Florida}
\author{A.~Norrick}                       \affiliation{\WM}
\author{A.~Olivier}                       \affiliation{\Rochester}
\author{V.~Paolone}                       \affiliation{\Pittsburgh}
\author{J.~Park}                       \affiliation{\Rochester}
\author{G.N.~Perdue}                      \affiliation{\FNAL}  \affiliation{\Rochester}
\author{M.A.~Ram\'{i}rez}                 \affiliation{\Guanajuato}
\author{R.D.~Ransome}                     \affiliation{\Rutgers}
\author{H.~Ray}                           \affiliation{\Florida}
\author{D.~Rimal}                         \affiliation{\Florida}
\author{P.A.~Rodrigues}                   \affiliation{\umiss}  \affiliation{\Rochester}
\author{D.~Ruterbories}                   \affiliation{\Rochester}
\author{H.~Schellman}                     \affiliation{\OregonState}  \affiliation{\Northwestern}
\author{C.J.~Solano~Salinas}              \affiliation{\UNI}
\author{H.~Su}                            \affiliation{\Pittsburgh}
\author{M.~Sultana}                       \affiliation{\Rochester}
\author{V.S.~Syrotenko}                   \affiliation{\Tufts}
\author{B.~Yaeggy}                        \affiliation{\USM}
\author{L.~Zazueta}                       \affiliation{\WM}

\collaboration{The MINER$\nu$A Collaboration}\ \noaffiliation

\date{\today}

\begin{abstract}
Elastic neutrino scattering on electrons is a precisely-known purely leptonic process that provides a standard candle for measuring neutrino flux in conventional neutrino beams.    Using a total sample of 810 neutrino-electron scatters after background subtraction, the measurement reduces the normalization uncertainty on the $\nu_\mu$ NuMI flux between 2 and 20 GeV from 7.6\% to 3.9\%.  This is the most precise measurement of neutrino-electron scattering to date, will reduce uncertainties on \minerva's absolute cross section measurements, and demonstrates a technique that can be used in future neutrino beams such as LBNF. 
\end{abstract}

\ifnum\sizecheck=0  
\maketitle
\fi


\section{Introduction}

Conventional neutrino beams are used extensively in the study of neutrino oscillations.  They provide intense sources to current experiments such as T2K~\cite{Abe:2017bay}, NOvA~\cite{Adamson:2017gxd}, and MicroBooNE~\cite{Acciarri:2016smi}, and planned experiments such as DUNE~\cite{Acciarri:2015uup} and T2HK~\cite{Abe:2018uyc}.  Produced by colliding a high energy proton beam on a solid target and focusing the resulting hadrons through one or more focusing horns, conventional neutrino beams carry large uncertainties in estimates of the total number of neutrinos in the beam as well as in their flavor and energy composition.  These uncertainties arise primarily from underlying uncertainties in the number and kinematic distributions of hadrons produced off the target, as well as in parameters describing beamline alignment and focusing.  

External measurements of hadron production off thin and thick targets have recently improved neutrino flux predictions substantially, but remaining uncertainties of order 10\%~\cite{Abe:2014nox,Aliaga:2016oaz} will begin affecting neutrino oscillation measurements in the next decade as oscillation measurements become dominated by systematic uncertainties.  Neutrino flux uncertainties are already the limiting uncertainty for measurements of neutrino interaction cross section measurements, which are themselves significant sources of systematic uncertainty in neutrino oscillation measurements~\cite{Abe:2018wpn,NOvA:2018gge}.

Measurements of the neutrino flux in near detectors require a standard-candle process with a known cross section.  Significant uncertainties in neutrino-nucleus scattering cross sections make these processes poor standard candle options.  Neutrino-nucleus interactions with low hadronic recoil energy (sometimes called "low-nu" interactions) have a cross section that is approximately constant with neutrino energy, and therefore can be used to measure the shape of the neutrino flux.

Neutrino scattering on electrons is a precisely calculable electroweak  process, although its cross section is small -- three orders of magnitude less than neutrino-nucleus scattering.  The \minerva~collaboration demonstrated~\cite{Park:2015eqa} use of this process as a constraint on the neutrino flux in the Low Energy (LE) configuration of the Neutrinos from the Main Injector (NuMI) beam at Fermilab.  This paper presents a similar measurement in the NuMI Medium Energy (ME) beam.   Increases in both total exposure and in neutrino flux per proton-on-target in the ME configuration provide a factor of nine increase in statistics over \minerva's earlier LE measurement.  Improvements in systematic uncertainties arising from better knowledge of neutrino-nucleus scattering and improved understanding of the electron energy scale in the \minerva~ detector have also significantly reduced the total systematic uncertainty.  

This measurement provides a precise constraint on the normalization of the neutrino flux prediction at MINERvA, and is complimentary to flux measurements using low hadronic recoil energy (``low-nu") neutrino-nucleus interactions, which can be used to measure the shape of the neutrino flux as as a function of energy.  Measurements of the flux shape using low-nu events have been made by MINERvA for the LE beam~\cite{DeVan:2016rkm} and are on-going for the ME beam.     

This paper is organized as follows.  The NuMI neutrino beam and its simulation is discussed in Sec.~\ref{sec:numi}, the \minerva~ detector in Sec.~\ref{sec:minerva}, the neutrino-elastic scattering measurement in Sections~\ref{sec:event_selection}-\ref{sec:systematics}, and the use of this measurement to reduce uncertainties in the neutrino flux in Sec.~\ref{sec:flux}.  Conclusions are presented in Sec.~\ref{sec:conclusion}.

\section{NuMI Beamline and Simulation}
\label{sec:numi}
The NuMI neutrino beam, described in detail in Ref.~\cite{Adamson:2015dkw}, includes a 120-GeV primary proton beam, a two-interaction-length graphite target, two parabolic focusing horns, and a 675-m decay pipe.   The data discussed here were taken between September 2013 and February 2017, when the beam was in the ME configuration optimized for the NOvA off-axis experiment.  In this configuration, the target begins 194 cm upstream of the start of the first focusing horn, creating a higher energy neutrino beam than was used for \minerva's earlier measurement~\cite{Park:2015eqa} of neutrino-electron scattering.   The current in the horns can be configured to focus positively- or negatively-charged particles, resulting in neutrino- or antineutrino-enhanced beams, respectively.  For the data discussed here, the current had an amplitude of 200 kA and was oriented to create a neutrino-enhanced beam.  The dataset corresponds to $1.16\times10^{21}$ protons on target.   

\begin{figure}[ht]
\centering
  \includegraphics[width=0.5\textwidth]{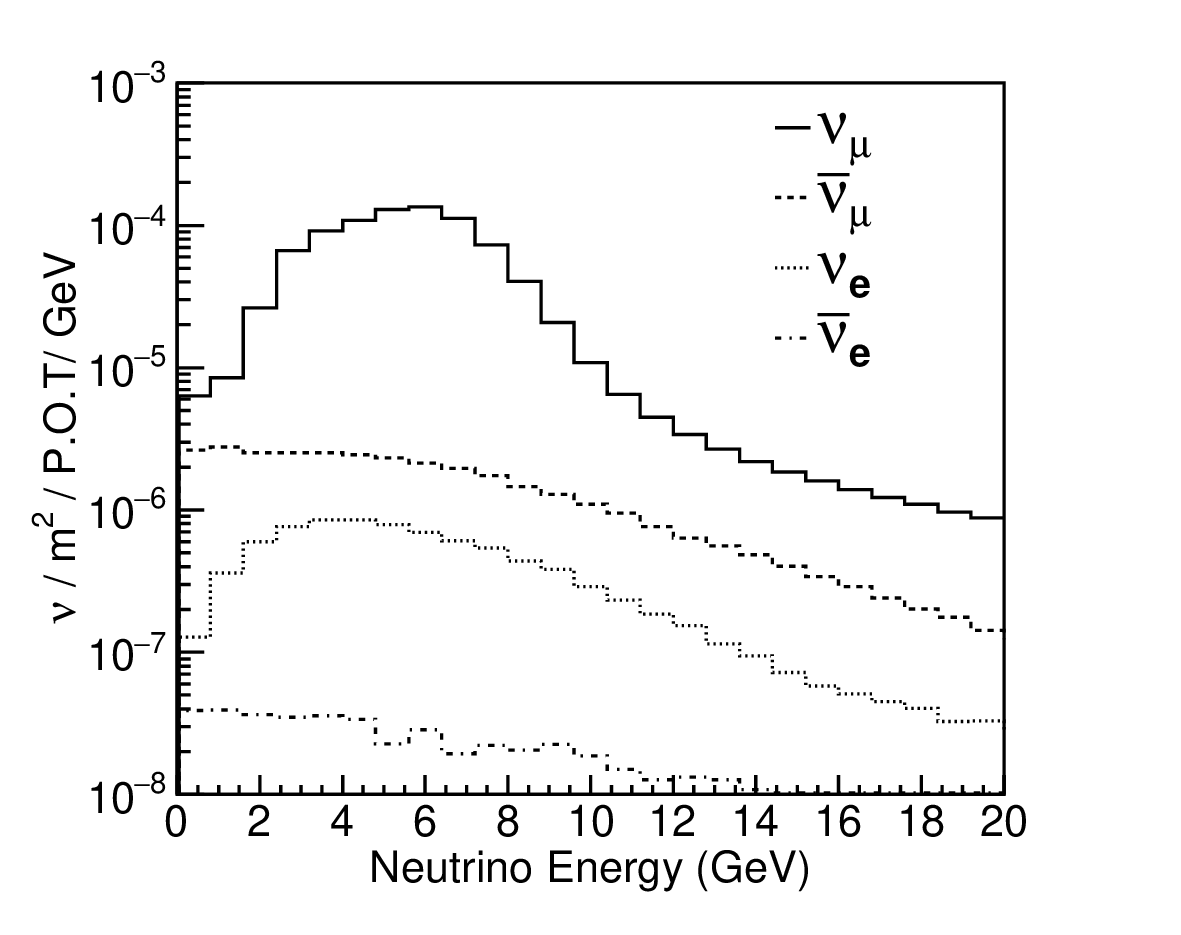} 
\caption{Simulated $\nu_\mu$, $\bar{\nu}_\mu$, $\nu_e$, and $\bar{\nu}_e$ fluxes at \minerva~versus neutrino energy in the ME neutrino-mode configuration of the NuMI beam.  }
\label{fig:flux_species}
\end{figure}

NuMI is simulated using a \gf-based model of the NuMI beamline\footnote{NuMI experiments share a common simulation of the NuMI beam known as g4numi; g4numi version 6 built against \minerva~version 4.9.2.p3 with the FTFP\_BERT physics list is used here.}.   There are known discrepancies between \gf predictions of proton-on-carbon and other interactions relevant to NuMI flux predictions.  \minerva~has developed a procedure for correcting \gf flux predictions using hadron production data~\cite{Aliaga:2016oaz}. Neutrino flux predictions after these corrections for the NuMI ME neutrino-mode configuration are shown in Fig.~\ref{fig:flux_species}.

\section{MINERvA Experiment and Simulation}
\label{sec:minerva}
The \minerva~ detector, described in detail in Ref.~\cite{Aliaga:2013uqz}, is composed of 208 hexagonal planes of plastic scintillator stacked along the z axis, which is nearly parallel to the NuMI beam axis\footnote{The beam axis (oriented so as to direct the NuMI beam through the earth to Minnesota) points 58 mrad downward compared to the detector z axis, which is parallel to the floor of the MINOS near detector hall.}.  Each plane is composed of 127 interleaved triangular strips of scintillator, arrayed in one of three orientations ($0^\circ$ and $\pm60^\circ$ with respect to the vertical) to facilitate three-dimensional particle reconstruction.  

The last ten planes at the downstream end of the detector are interspersed with 26-mm thick planes of steel, functioning as a hadronic calorimeter (HCAL).  The twenty planes upstream of the HCAL are separated by 2-mm thick sheets of lead, forming an electromagnetic calorimeter.  A 2-mm thick lead collar covers the outermost 15 cm of all scintillator planes (the side ECAL), and the planes are supported by steel frames embedded with scintillator (the side HCAL).  The upstream end of the detector also contains five planes of passive targets constructed of lead, iron, and carbon, as well as water and cryogenic helium targets.  The MINOS near detector sits 2 m downstream of \minerva~ but is not used in this analysis.  Only energy depositions in the inner tracker, ECAL, and HCAL are used in the measurement in this analysis.  

The energy depositions in the scintillator strips are read out through wavelength-shifting fibers to multi-anode PMTs using a data aquisition system described in Ref.~\cite{Perdue:2012hg}.  Calibration of the detector, described in Ref.~\cite{Aliaga:2013uqz}, relies on through-going muons created in neutrino interactions upstream of the \minerva~ detector.  The timing resolution of individual hits is better than 4 ns.  The electromagnetic energy scale is cross-checked with a sample of electrons originating from muon decays and also with a sample of $\pi^0\rightarrow\gamma\gamma$  candidates, discussed further in Sec.~\ref{sec:systematics}.

Neutrino interactions in the \minerva~ detector are simulated using the GENIE neutrino event generator~\cite{Andreopoulos:2009rq}\cite{Andreopoulos:2015wxa} version 2.12.6.  This version of GENIE includes a tree-level calculation of the neutrino-electron elastic scattering cross section.  This cross section is modified to account for modern calculations of electroweak couplings~\cite{Erler:2013xha} and for first-order radiative corrections~\cite{Bardin:1983yb}, as described in the Appendix.   Quasi-elastic neutrino-nucleus interactions are simulated with a Relativistic Fermi Gas model~\cite{SMITH1972605} and the Llewellyn-Smith formalism~\cite{LlewellynSmith:1971uhs}; \minerva~modifies the quasi-elastic model with weak charge screening (RPA) corrections~\cite{Nieves:2004wx} as described in Ref.~\cite{Gran:2017psn}.  Although not part of the GENIE default model, \minerva~adds a simulation of quasi-elastic-like interactions off correlated nucleon pairs using the Valencia IFIC model~\cite{PhysRevC.83.045501}.  

Models of Rein and Seghal are used for both neutrino-nucleon interactions with resonance production~\cite{REIN198179} and coherent pion production~\cite{REIN198329}.  \minerva~ adds a simulation of diffractive neutral-current $\pi^0$ production off hydrogen based on a model by Rein~\cite{Rein:1986cd} that is implemented in GENIE but not turned on by default.   Deep inelastic scattering is simulated using the Bodek-Yang Model~\cite{Bodek:2002ps}. Intranuclear rescattering is simulated using the GENIE INTRANUKE-hA package.  

Propagation of particles through the \minerva~detector is modeled with a simulation based on \gf version 4.9.4.p02 with the QGSP\_BERT physics list.  Overlapping activity and dead time from other neutrino interactions is simulated by overlaying \minerva~data beam spills on simulated events.  NuMI spills are delivered in six bunches spanning a total of 10 $\mu$s.  While the data considered here were taken, the intensity of NuMI bunches varied between $1\times10^{12}$ and $9\times10^{12}$ protons.  \minerva's simulation of accidental activity was upgraded for the ME era to ensure that the in-spill timing of simulated events and intensity of unrelated coincident energy depositions precisely matches those distributions in the data.  

\section{Event Reconstruction and Selection}
\label{sec:event_selection}

Neutrino-electron elastic scatters in the \minerva~ detector typically appear as single electron showers nearly parallel to the neutrino beam direction.  Reconstruction of events consistent with this topology begins by separating hits in each 10 $\mu$s NuMI spill in time, producing ``time slices," consistent with a single neutrino interaction.  Hits within a single time slice on adjacent strips within each plane are then grouped into clusters.  Electron shower candidates are then identified via a cone algorithm that is seeded by one of two types of objects: tracks or groups of clusters.  In most cases, clusters are formed into tracks as described in Ref.~\cite{Aliaga:2013uqz} and fit using a Kalman filter that provides a start vertex and track direction. The radiation length of scintillator corresponds to approximately 25 \minerva~ planes for a track parallel to the detector $z$ axis.  Occasionally, when an electron begins to shower after traversing a small number of
planes, a track cannot be reconstructed using the Kalman fitter method. In
such cases spatially continuous clusters that are not consistent with a track
are grouped together and fit with a chi-square minimization method to assign a vertex and angle
to the group of clusters.

\begin{figure}[ht]
\centering
  \includegraphics[width=0.5\textwidth]{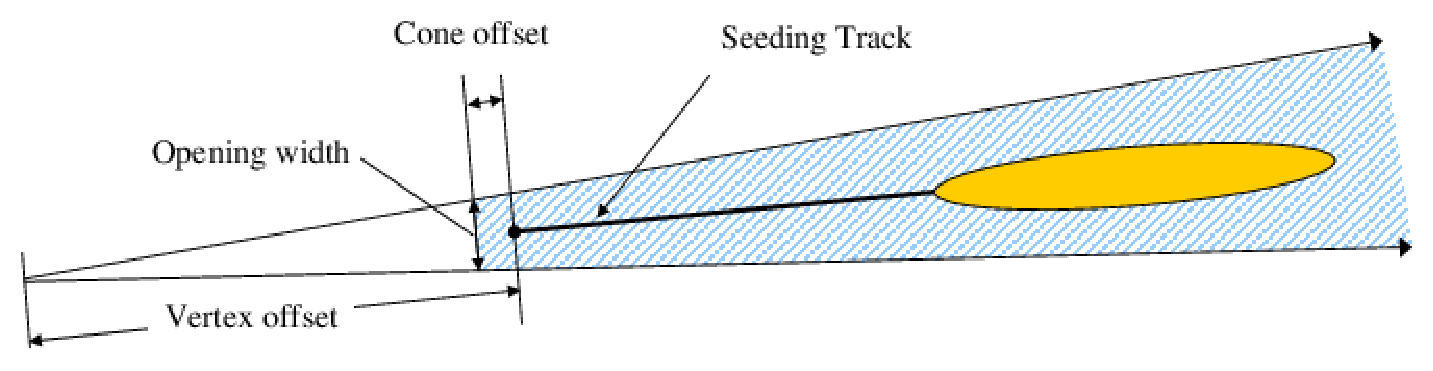}
\caption{Illustration of the cone used to identify energy depositions that are part of the electron shower.  For this analysis, the cone offset is 50 mm, the opening width is 80 mm and cone opening angle is $10^\circ$. }
\label{fig:cone}
\end{figure}

Given the vertex and angle of the reconstructed tracks or groups of clusters, a cone is formed that begins upstream of the vertex and extends downstream until no minimum-ionizing-level energy depositions are found within the cone volume.  The axis of the cone lies along the reconstructed direction of the electron candidate.  The cone has an opening angle of 10$^\circ$, is positioned such that the width of the cone 80 mm upstream of the vertex is 50 mm, and is truncated upstream of this 80-mm point (see Fig.~\ref{fig:cone}).    These parameters were chosen to maximize efficiency and containment for the signal in the simulation.

The calibrated energy of the clusters within the cone is summed, weighted by calorimetric constants obtained using the simulation that account for passive material in the detector as well as the overall difference between electron energy and energy deposited in \minerva.  The measurement of the electron's energy (angle) has a resolution ranging from 60 MeV (0.7$^\circ$) in the lowest energy bin reported here (0.8-2 GeV) to 40 MeV (0.3$^\circ$) for the highest energies (above 9 GeV).

Only electrons within \minerva's main tracker volume are considered.  Specifically, the transverse position of the electron's vertex must lie within a hexagon with apothem 81.125 cm and the vertex longitudinal position must lie within the central 111 planes of the \minerva~ tracker.  This corresponds to a fiducial mass of 5.99 metric tons.     

A series of cuts is applied to the electron candidates to ensure well-reconstructed electrons.  These cuts, described in the following paragraphs, as well as the background-reduction cuts described later are identical to those used in the earlier LE analysis and they are described in further detail  in Ref.~\cite{Park:2015eqa}.  

To ensure the electron candidate is contained within the detector, showers ending in the final four planes of the \minerva~ HCAL are rejected, as are those with an endpoint within 2 cm of the edge of the detector in any of the three views.  Events originating in the upstream nuclear target region of \minerva~ are vetoed by summing the visible energy inside a cylinder with a radius of 30 cm and with an axis that extends along the upstream extrapolation of the electron candidate.  Events with more than 300 MeV of energy within this cylinder are rejected. 

All of the electron candidates here arise from events in the \minerva~ tracker; those with energy deposited in the HCAL must traverse the entirety of the ECAL, and are expected to lose most of their energy in the ECAL.  To reduce hadronic backgrounds, at least 80\% of the total energy in the ECAL + HCAL must be deposited in the ECAL.      

When an event occurs in the \minerva~ detector, activated electronics channels are not able to detect a subsequent event for approximately 150 ns after the initial event.  This dead time has a particularly adverse effect on event vertex reconstruction, as dead channels can cause events originating outside of the fiducial volume to have reconstructed vertices inside the fiducial volume.  The data and simulation save a record of when each channel is in a dead state.  To eliminate events where the vertex has been falsely placed due to dead time, the axis of the electron shower is extrapolated four planes upstream of the reconstructed vertex.  If two or more of the intersected strips or their immediate neighbors are dead, the event is rejected.  

Additional cuts are made to reduce backgrounds from neutrino-nucleus interactions.  Only electron candidates above 0.8 GeV are considered in order to reduce backgrounds from neutral-current $\pi^0$ production.  Neutral pion backgrounds are further reduced by identifying events with two energy depositions in the ECAL with transverse separation consistent with two nearly colinear photons resulting from $\pi^0$ decay.  

\begin{figure}[ht]
\centering
  \includegraphics[width=0.5\textwidth]{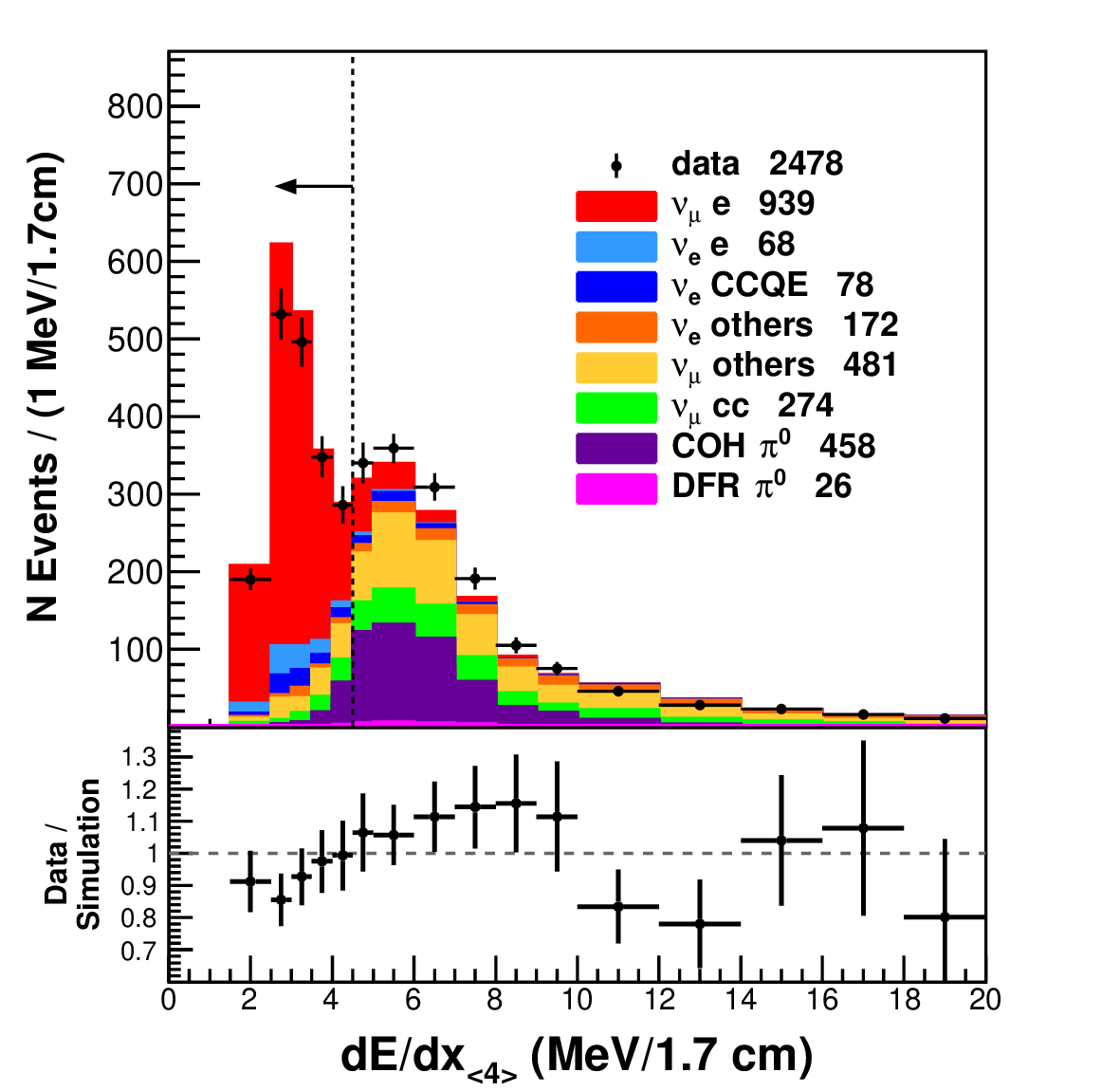}
\caption{Average energy deposition in the first four planes of the electron candidate for events passing all other cuts in data and simulation (above) and the ratio of data to simulation (below).  The error bars on the data points include statistical uncertainties only.  The error bars on the ratio include both statistical uncertainties in data and statistical and systematic (see Sec.~\ref{sec:systematics}) uncertainties  in the simulation.  Backgrounds have been tuned using the procedure described in Sec.~\ref{sec:background}.  }
\label{fig:dedx}
\end{figure}

One of the most effective cuts is on the average energy deposition of the electron in the first four planes of the shower ($\mathrm{dE/dx}_{\langle4\rangle}$), in which energy deposited by converted photons ($e^+e^-$ pairs), which often come from  neutral pions, is approximately twice that of an electron.  The $\mathrm{dE/dx}_{\langle4\rangle}$ of events passing all other cuts is shown in Fig.~\ref{fig:dedx}.  This quantity is required to be less than 4.5 MeV/1.7 cm.  

\begin{figure}[ht]
\centering
  \includegraphics[width=0.5\textwidth]{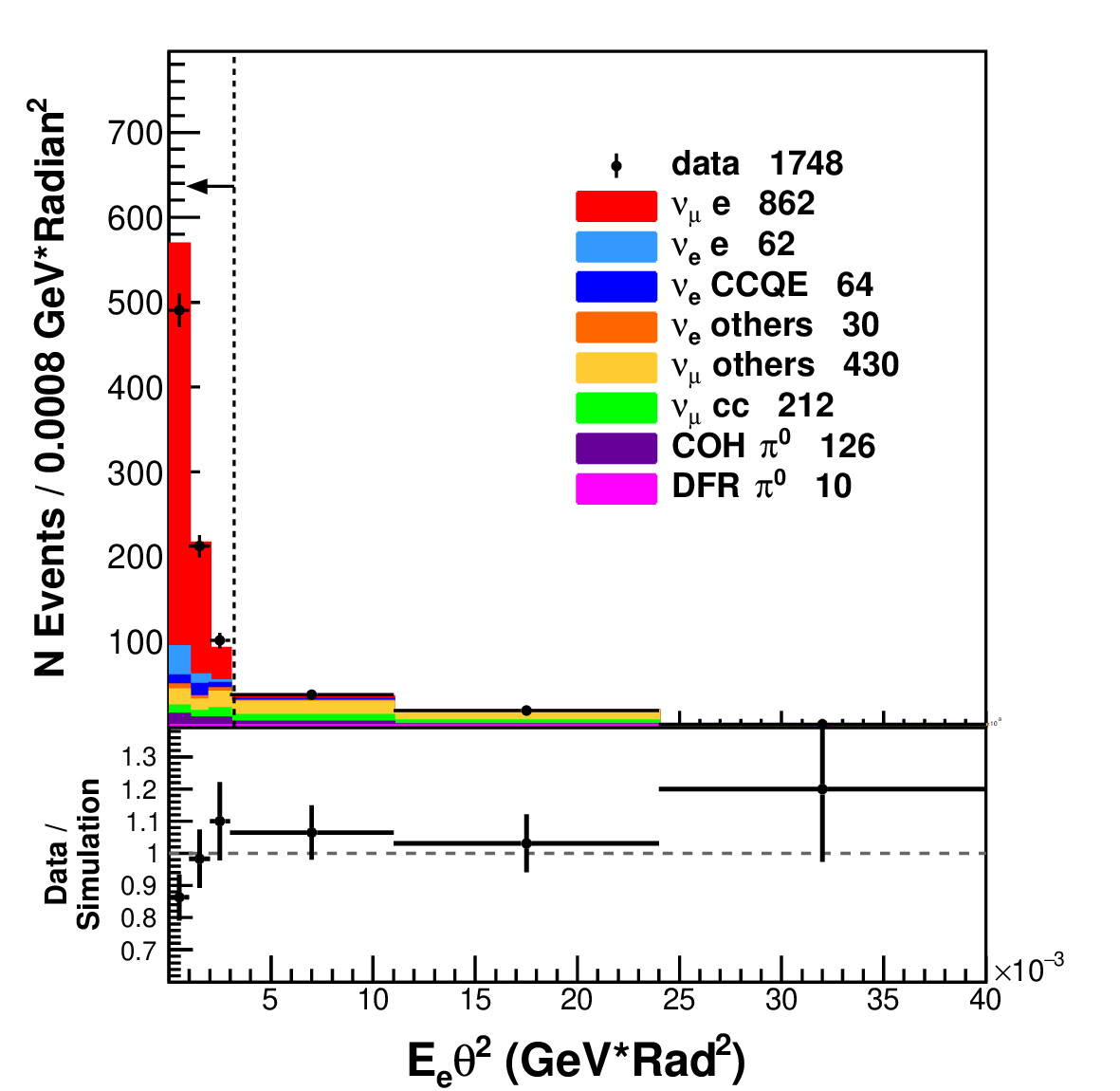} 
\caption{The electron energy times the square of the electron angle with respect to the beam for candidate events passing all other cuts in data and simulation (above) and the ratio of data to simulation (below).  The error bars on the data points include statistical uncertainties only.  The error bars on the ratio include both statistical uncertainties in data and statistical and systematic (see Sec.~\ref{sec:systematics}) uncertainties in the simulation.  Signal events are required to have $E_e\theta^2 < 0.0032$ GeV radian$^2$.  Backgrounds have been tuned using the procedure described in Sec.~\ref{sec:background}. }
\label{fig:eth2}
\end{figure}

To ensure that there are no visible particles in the event other than the electron, energy outside of the shower reconstruction cone but within 5 cm of the cone's outer shell is summed and required to be less than 120 MeV for electron candidates with energy below 7 GeV or less than  $7.8 E_e + 65.2$ MeV otherwise, where $E_e$ is the reconstructed energy of the electron in GeV.  

Overlapping particles may also be reconstructed as part of the electron shower itself.  The upstream portion of these events is typically wider than that of a single electron; electron candidates are therefore required to have an transverse residual RMS less than 20 mm in the upstream third of the shower. A cut is further made on the transverse residual RMS over the full shower,  calculated separately for X, U and V-view clusters.  The maximum of these three RMS values is required to be less than 65 mm.  The longitudinal energy profile of the shower is also required to be consistent with that of a single electromagnetic particle. 

Because \minerva~ detector planes are arrayed in an XUXV pattern, approximately 50\% of the electron shower's energy is deposited in X planes, and 25\% each in U and V planes.  This is not necessarily the case for showers involving multiple particles, which will usually overlap the electron candidate in some views and not in others.  To further reduce these events, two quantities are constructed:
\begin{eqnarray}
E_{XUV} &= \frac{E_X-E_U-E_V}{E_X+E_U+E_V} \\
E_{UV} &= \frac{E_U-E_V}{E_U+E_V},
\end{eqnarray}
and electron candidates are required to satisfy $E_{XUV}<0.28$ and $E_{UV}<0.5$.

High energy electron showers tend to follow a straight line through the \minerva~ detector, whereas interactions of hadronic particles will often cause hadron showers to appear bent.  To help eliminate hadronic-shower backgrounds, a bending angle is formed by defining two line segments, one from the start point of the shower to its midpoint, and one from the midpoint to the endpoint.  The angle between these lines is required to be less than $9^\circ$.  This and other background rejection criteria were determined by optimizing signal significance according to the simulation.    


After all of the cuts described above, the dominant background in the sample is $\nu_e$ and $\bar{\nu}_e$ quasi-elastic scattering ($\nu_e n \rightarrow e^-p$ and $\bar\nu_e p \rightarrow e^+ n$) in which the recoiling nucleon is not observed, which is typical for quasi-elastic events with low 4-momentum transfer squared ($Q^2$).  Although these categories have an identical final-state particle signature to neutrino-electron scattering, they can be substantially reduced with kinematic cuts.  One kinematic quantity that is useful here is the product of electron energy and the square of the angle of the electron  with respect to the neutrino beam ($E_e\theta^2$).  For neutrino-electron elastic scattering, $E_e\theta^2$ is kinematically constrained to be less than twice the electron mass.  The $E_e\theta^2$ distribution for events passing all other cuts described here is shown in Fig.~\ref{fig:eth2}.  Candidate events are required to have $E_e\theta^2<0.0032$ GeV rad$^2$.  

To further reduce quasi-elastic background events, $Q^2$ is reconstructed assuming a quasi-elastic hypothesis:
\begin{eqnarray}
Q^2 &= 2m_n\left(E_\nu-E_e\right) \\
E_\nu &= \frac{m_nE_e-m_e^2/2}{m_n-E_e+p_e\cos{\theta}},
\end{eqnarray}
where $m_n$ and $m_e$ are the masses of the neutron and electron, respectively, $p_e$ is the momentum of the electron, and $\theta$ is the angle of the electron with respect to the neutrino beam.  Candidate events are required to have $Q^2$ less than 0.02 GeV$^2$.  

\begin{figure}[ht]
\centering
  \includegraphics[width=0.5\textwidth]{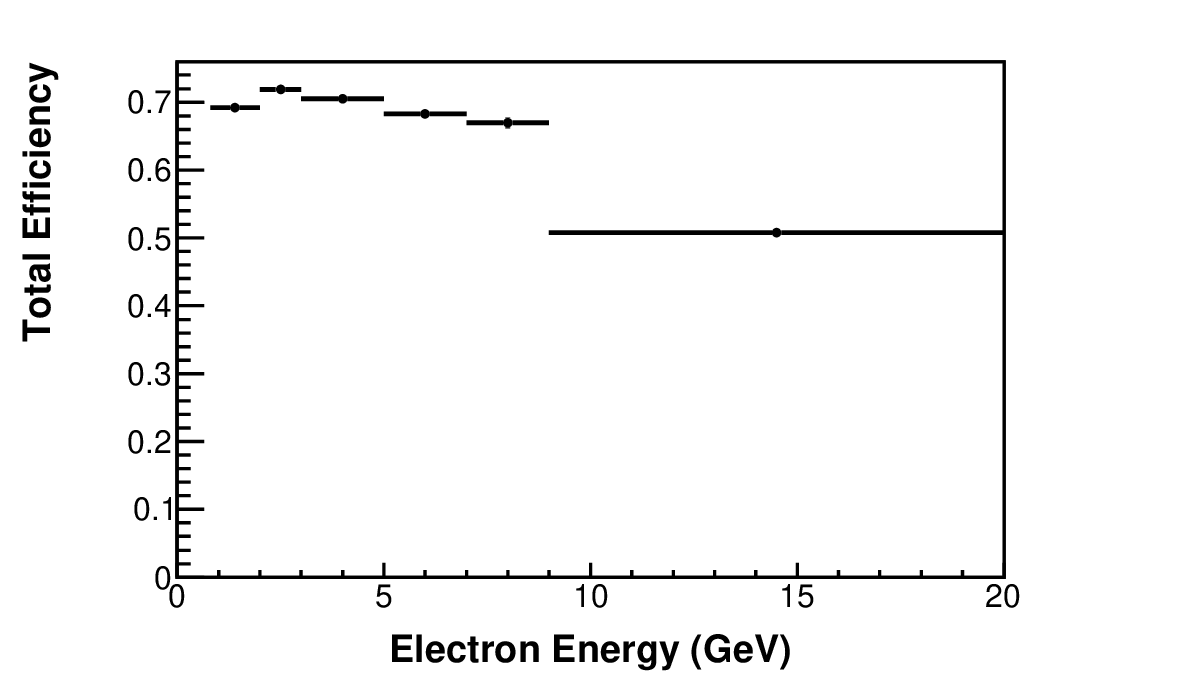} 
\caption{Total efficiency of neutrino-electron scattering candidates after all selection cuts.  }
\label{fig:efficiency}
\end{figure}

The signal efficiency of the event reconstruction selection after all cuts is shown in Fig.~\ref{fig:efficiency}.

\section{Background Subtraction}
\label{sec:background}
\begin{figure}[ht]
\centering
  \includegraphics[width=0.45\textwidth]{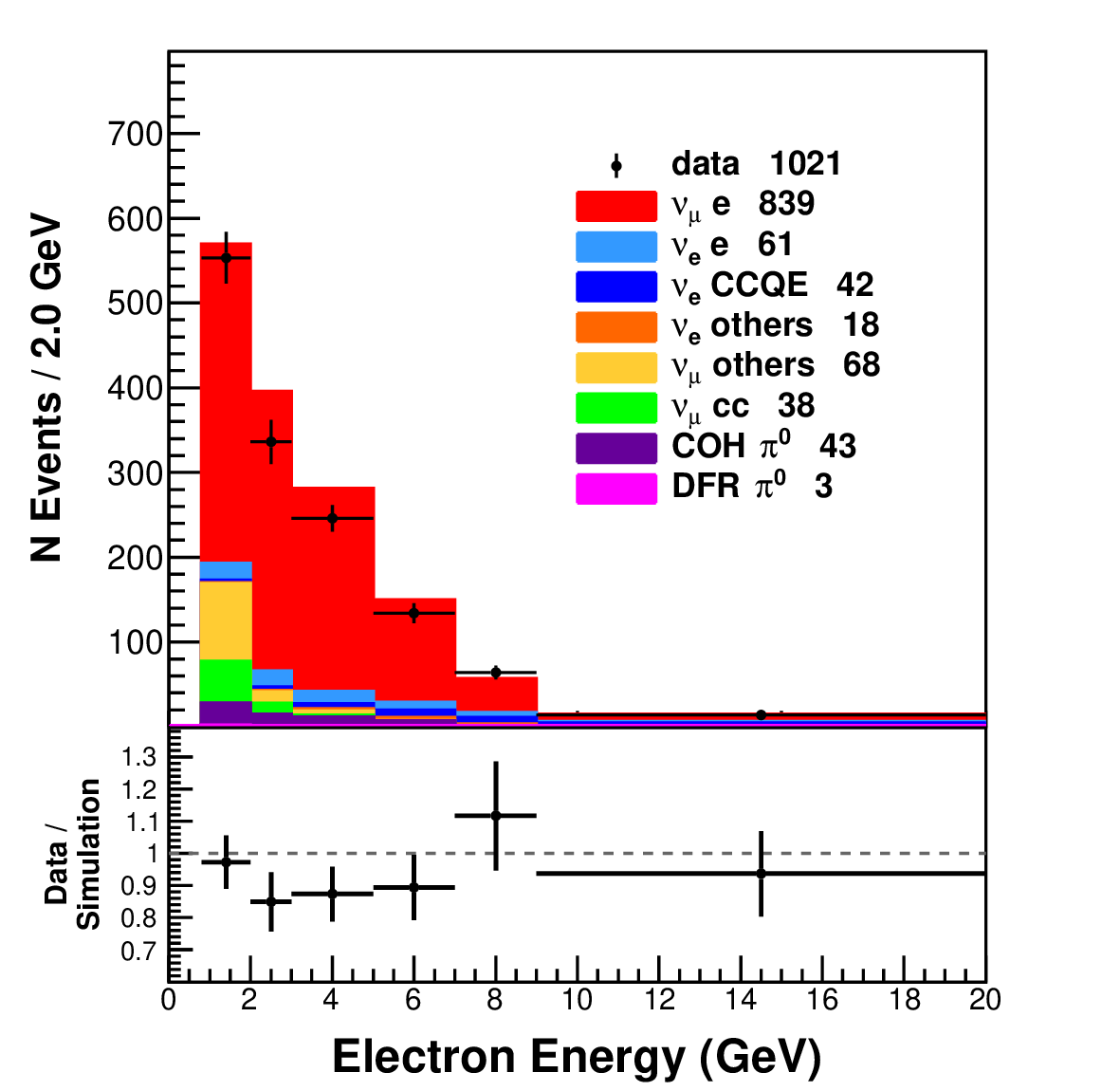} 
\caption{Reconstructed electron energy in events passing all cuts in data and simulation (above) and the ratio of data to simulation (below).  The error bars on the data points include statistical uncertainties only.  The error bars on the ratio include both statistical uncertainties in data and statistical and systematic (see Sec.~\ref{sec:systematics}) uncertainties  in the simulation. Backgrounds have been tuned using the procedure described in Sec.~\ref{sec:background}. The highest energy bin includes all events with $E_e>9$ GeV events, including events with $E_e>20$ GeV.}
\label{fig:energy}
\end{figure}
The energy distribution of electron candidates passing all cuts is shown in Fig.~\ref{fig:energy} for data and simulation.  Of the 1112 predicted candidates, 212 are background.  Nearly half of the predicted background events are from $\nu_\mu$ neutral-current interactions, including coherent and diffractive $\pi^0$ production. About 20\% arise from $\nu_\mu$ charged-current interactions, with the remainder due to $\nu_e$ interactions, primarily charged-current quasi-elastic scattering.  The backgrounds are largest at low reconstructed electron energies, where misreconstructed $\nu_\mu$ events are particularly prevalent, and at high reconstructed electron energies, where backgrounds from $\nu_e$ quasi-elastic scattering are large.   

The backgrounds predicted by the GENIE-based simulation have been constrained using four kinematic sidebands.  Sidebands 1-3 use different $E_e\theta^2$ and $\mathrm{dE/dx}_{\langle4\rangle}$ cuts than the signal region: $0.005 < E_e\theta^2 < 0.1$ GeV radian$^2$ and $\mathrm{dE/dx}_{\langle4\rangle} < 20$ MeV/1.7 cm.   To enhance statistics in the sidebands, the cuts on shower mean radius in the first third of the shower and on $Q^2_{\mathrm CCQE}$ are removed.  Sideband 1 further requires that the minimum dE/dx in the second through sixth planes be greater than 3 MeV/1.7 cm.  Sideband 2 (3) events are required to have minimum dE/dx in the second through sixth planes less than 3 MeV/1.7 cm and have electron energy less than (greater than) 1.2 GeV.    Sideband 4 satisfies all of the main analysis cuts except that $\mathrm{dE/dx}_{\langle4\rangle}$ is required to be between 4.5 and 10 MeV/1.7 cm.  

The sidebands are designed to constrain four categories of background:  1) neutral-current coherent $\pi^0$ production, 2) charged-current $\nu_\mu$ interactions, 3) neutral-current $\nu_\mu$ interactions (excluding diffractive and coherent $\pi^0$ production), and 4) $\nu_e$ interactions. 
Sideband 1 is approximately 30\% $\nu_\mu$ charged-current interactions, 50\% $\nu_\mu$ neutral-current interactions (excluding diffractive and coherent $\pi^0$ production), 10\% coherent $\pi^0$ production and 10\% $\nu_e$ interactions.  Sideband 2 is composed of approximately one third $\nu_\mu$ interactions and two thirds non-diffractive or coherent $\nu_\mu$ neutral-current interactions.  Sideband 3 is approximately 50\% $\nu_e$ interactions, with the remaining half split roughly evenly between $\nu_\mu$ charged-current and non-diffractive, non-coherent $\nu_\mu$ neutral-current interactions.  

Prior to background constraint, there is an excess in data in Sideband 4, the high $\mathrm{dE/dx}_{\langle4\rangle}$  sideband.  This sideband is populated by all of the background sources discussed above except $\nu_e$ interactions, and according to the simulation it consists  primarily of events with $\pi^0$s in the final state.    
A similar excess was seen in a separate \minerva~ measurement of $\nu_e$ quasi-elastic-like scattering~\cite{Wolcott1}, and it was found to be consistent with neutral-current diffractive $\pi^0$ production~\cite{Wolcott2}.  The GENIE model for neutral-current diffractive scattering used here predicts very few events in the signal or sideband regions of this analysis, but significant contributions from similar coherent $\pi^0$ production\footnote{GENIE does not currently contain a model of coherent photon production, but this process may also be present and would appear similar to coherent $\pi^0$ production background events in the MINERvA detector.}.   The excess in sideband 4 is attributed to coherent events, allowing the normalization of that background to float in the background fits, which are performed by computing a $\chi^2$ summed over distributions in each of the four sidebands.  Because \minerva~ studies of both neutral-current diffractive~\cite{Wolcott2} and charged-current coherent $\pi^0$ production~\cite{Mislivec:2017qfz} have found significant discrepancies with GENIE predictions that vary with energy, the normalization of the coherent background is allowed to vary separately for each of the six electron energy bins.  For the other three backgrounds, the fit includes a single normalization factor that is constant with reconstructed energy.  The best fit normalizations of each of the floated background components is shown in Table~\ref{tab:backgrounds}.

\begin{table}[ht]
\centering
\begin{tabular}{ c | c }
\hline \hline
Process & Normalization \\ \hline
$\nu_e$ &  $0.87\pm0.03$ \\
$\nu_\mu$ CC & $1.08\pm0.04$ \\
$\nu_\mu$ NC & $0.86\pm0.04$ \\
NC COH $0.8 < E_e < 2.0$ GeV & $0.9 \pm 0.2$\\
NC COH $2.0 < E_e < 3.0$ GeV & $1.0 \pm 0.3$ \\
NC COH $3.0 < E_e < 5.0$ GeV & $1.3 \pm 0.2$ \\
NC COH $5.0 < E_e < 7.0$ GeV & $1.5 \pm 0.3$\\
NC COH $7.0 < E_e < 9.0$ GeV & $1.7 \pm 0.8$ \\
NC COH $9.0 < E_e$ & $3.0 \pm 0.9$ \\
\hline \hline
\end{tabular}
\caption{Background normalization scale factors extracted from the fits to kinematic sidebands, with statistical uncertainties.  }
\label{tab:backgrounds}
\end{table}

\begin{figure}[ht]
\centering
  \includegraphics[width=0.45\textwidth]{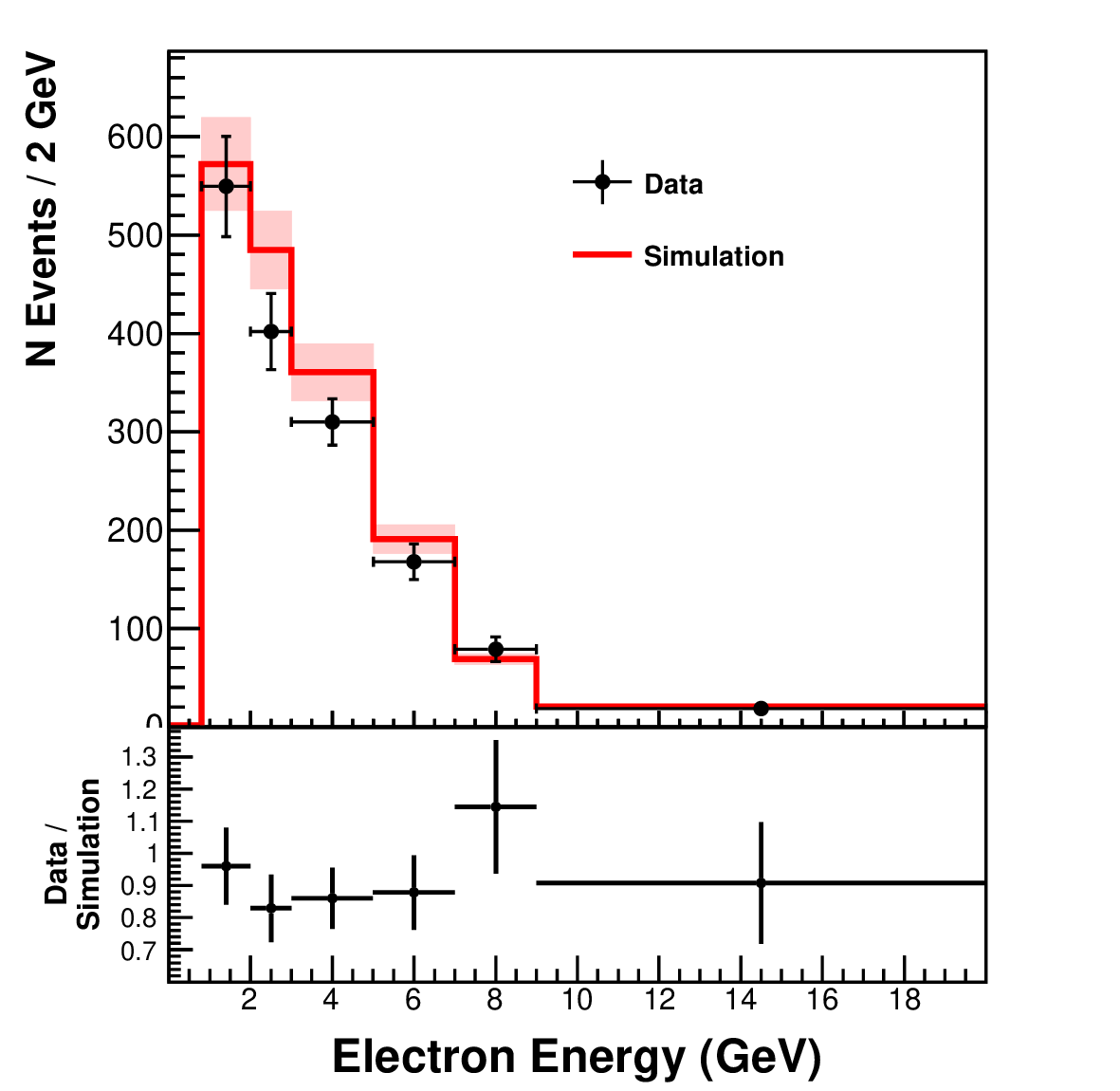} 
\caption{Reconstructed electron energy after background subtraction and efficiency correction in data and simulation (above) and the ratio of data to simulation (below).  The data error bars include both statistical and systematic uncertainties, as described in  Sec.~\ref{sec:systematics}.  The simulated spectrum error bars include both simulated statistics and neutrino flux uncertainties.  The ratio errors combine all of the uncertainties plotted in the top panel.  The highest energy bin includes all events with $E_e>9$ GeV events, including events with $E_e>20$ GeV.}
\label{fig:effcor}
\end{figure}

To obtain a background-subtracted electron energy spectrum in data, backgrounds predicted by the simulation are scaled by the factors given in Table~\ref{tab:backgrounds} and subtracted from the electron energy spectrum in data as shown in Fig.~\ref{fig:energy}.  This spectrum is then corrected using the efficiency shown in Fig.~\ref{fig:efficiency}.  The electron energy spectra in the data and the simulation after background subtraction and efficiency correction are shown in Fig.~\ref{fig:effcor}.   

\section{Systematic Uncertainties}
\label{sec:systematics}

\begin{figure}[ht]
\centering
  \includegraphics[width=0.45\textwidth]{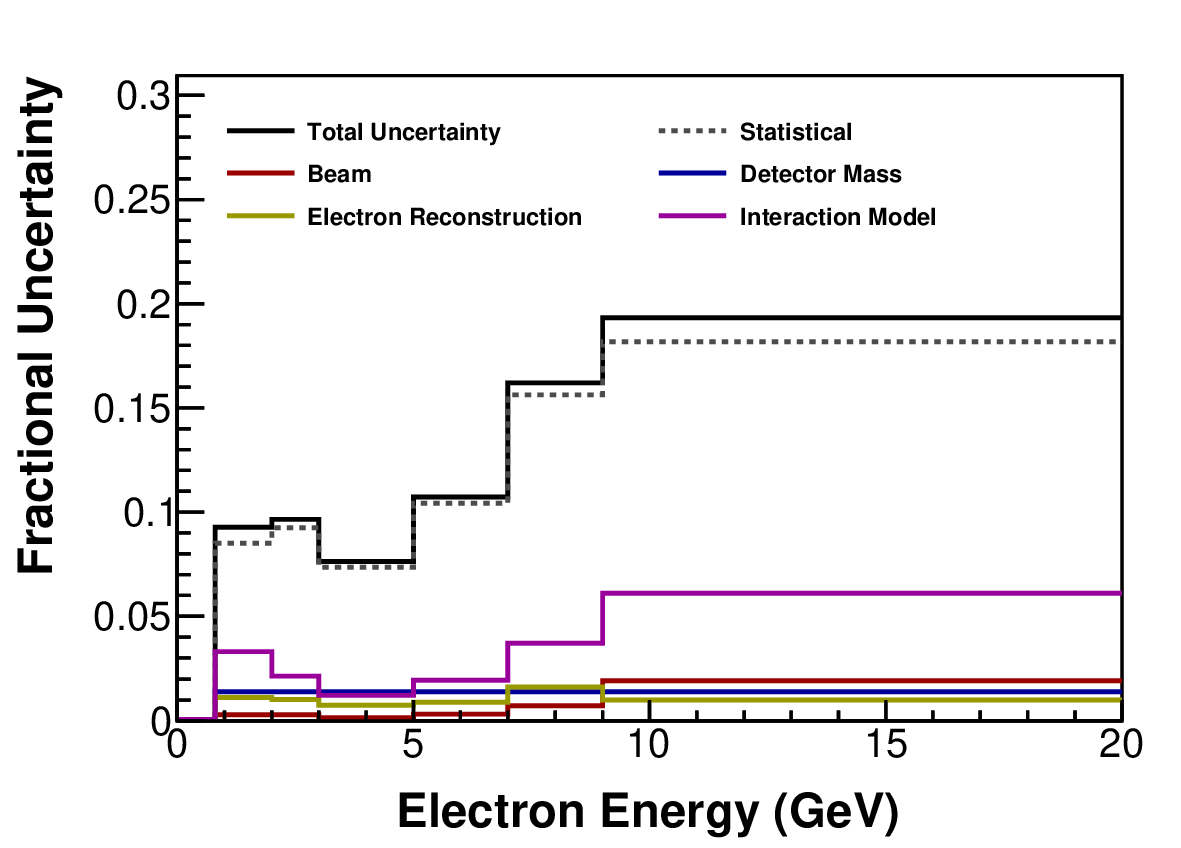} 
\caption{Summary of fractional systematic uncertainties on the background subtracted, efficiency corrected distributions.}
\label{fig:uncertainty}
\end{figure}

The background-subtracted, efficiency-corrected distribution shown in Fig.~\ref{fig:effcor} forms the basis of the flux constraint described in the Sec.~\ref{sec:flux}.  This distribution is subject to a variety of systematic uncertainties, which are summarized in Fig.~\ref{fig:uncertainty} and Table~\ref{tab:error_summary}.  The distribution, uncertainties and covariance matrix are also available in Table~\ref{tab:results}.  These are evaluated by identifying underlying uncertain parameters in the simulation, shifting those parameters by their uncertainty, and performing the analysis (including background subtraction and efficiency correction) with the shifted simulation.  The resulting change in the background subtracted, efficiency corrected spectrum is used to form a covariance matrix that encapsulates the systematic uncertainties due to that parameter and their correlations.  In some cases, it is appropriate to shift a parameter by +1 and -1 sigma, which produces two covariance matrices.  These covariance matrices are averaged to estimate the covariance of a distribution due to the parameter in question.  In the case of the neutrino flux uncertainties, there are many underlying uncertain parameters that are highly correlated with one another.  In this case, the many universes method is used, wherein many simulations are created, with each of the flux parameters pulled randomly from their probability distributions.  The total flux covariance matrix is formed from the average of the covariance matrix obtained with each simulation.  

There are several systematic uncertainties associated with electron reconstruction, such as the electromagnetic energy scale of the \minerva~ detector.  Uncertainty on the energy scale in the tracker and electromagnetic calorimeter was estimated by comparing energy of reconstructed $\pi^0$ candidates in charged-current $\nu_\mu$ events between data and simulation.  This comparison indicated that the tracker energy scale was well-modeled in the simulation, and this conclusion was supported by data-simulation comparisons of the spectra of low energy electrons from stopped muon decays.  The $\pi^0$ sample indicated a 5.8\% mismodeling of the energy scale in the electromagnetic calorimeter.  Energy deposits in the calorimeter were adjusted by 5.8\% and an overall uncertainty of 1.5\% in the electromagnetic response of the tracker and electromagnetic calorimeter was applied, based on the precision of the $\pi^0$ sample.  A conservative 5\% uncertainty on the energy scale in the hadronic calorimeter was assumed, based on a small sample of electrons reconstructed in the \minerva~test beam detector~\cite{testbeam,joshthesis}.  These energy scale uncertainties result in a small (0.1\%) uncertainty on the measured number of neutrino-electron scatters.    

\subsection{Electron Reconstruction Uncertainties}

In the previous \minerva~ measurement of this channel~\cite{Park:2015eqa}, one of the largest systematic uncertainties was due to the electron reconstruction efficiency.  That uncertainty was estimated from a study of muons reconstructed in the MINOS near detector that were projected backwards into \minerva, which found a 2.7\% difference between efficiencies in data and simulation due to accidental NuMI beam activity.  Improvements in the simulation of accidental activity have reduced that difference to 0.4\% for this analysis.  Additionally, a visual scan of event displays of electrons that failed reconstruction in the simulation was performed for this analysis.  Most of these failures were caused by accidental activity, but a small (0.4\%) fraction of electrons were misreconstructed for reasons that could not be discerned and were unrelated to accidental activity. A conservative 100\% uncertainty is assigned to these events, resulting in a total $0.4 \bigoplus 0.4 = 0.57$\% uncertainty on electron reconstruction efficiency, which in turn becomes a 0.57\% uncertainty on the neutrino-electron scattering rate.  

\subsection{Beam Uncertainties}

Small uncertainties in both the background estimation and efficiency estimation arise from sources related to the NuMI beam.  Uncertainties in the NuMI neutrino energy spectra arise primarily from hadron production and beam alignment.  These are estimated using the same procedure used for the LE configuration of the NuMI beam~\cite{Aliaga:2016oaz}.  Uncertainties in the $\nu_\mu$ flux range from 7-12\% depending on energy, and result in a 0.2\% uncertainty in the measured neutrino-electron scattering rate, primarily through the background subtraction procedure.  
Uncertainty in the angle of the NuMI beam is estimated by comparing muon angular spectra in charged-current $\nu_\mu$ candidates with low hadron recoil in data and simulation.  This results in a 0.5 mrad uncertainty in the beam angle, leading to a  0.1\% uncertainty on the neutrino-electron scattering rate.

\subsection{Interaction Model Uncertainties}

The largest category of systematic uncertainty is that associated with the neutrino interaction models used in the simulations.  These are largely assessed using the model parameter variations that are provided as event weights in the GENIE event generator.  Several uncertainties are also added in addition to those provided by the GENIE developers, as described below.  

\begin{figure*}[ht]
\centering
  \includegraphics[width=0.9\textwidth]{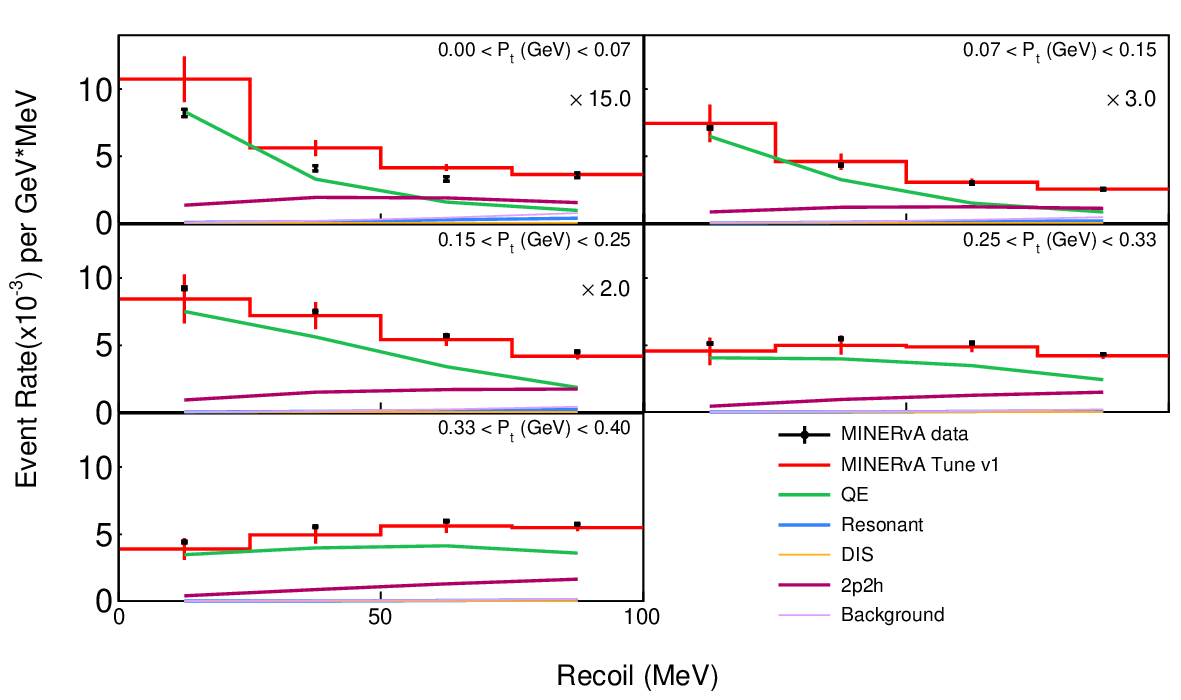}
\caption{Recoil energy distributions in bins of muon transverse momentum ($P_t$) in a sample of $\nu_\mu$ charged-current quasi-elastic scattering candidates from an ongoing measurement of quasi-elastic-like events in a $P_t$, $P_{||}$, recoil phasespace .  The quasi-elastic backgrounds in this analysis are $\nu_e$ interactions with low $P_t$ and low recoil ($P_t<0.15$, top row).  The sideband used to constrain this background occupies moderate $P_t$ and low recoil ($0.15<P_t<0.4$, second and third rows).  The error bars on the data are statistical; the simulation error bars include both statistical uncertainties and uncertainties associated with the RPA correction.  }  
\label{fig:ccqe}
\end{figure*}

Electron neutrino charged-current quasi-elastic scattering at low $Q^2$ is a significant background.  The analysis is particularly sensitive to uncertainty in the shape of the simulated $Q^2$ spectrum that is used to extrapolate backgrounds from the higher $Q^2$ sideband to low $Q^2$ signal region.   Uncertainties on the RPA correction to the quasi-elastic scattering model are taken from Ref.~\cite{Gran:2017psn}. A recent \minerva~ measurement of $\nu_\mu$ charged-current scattering indicates that even with this RPA correction, low $Q^2$ quasi-elastic events are overpredicted by approximately $30$\%~\cite{Ruterbories:2018gub}.  Further study of the hadronic recoil energy distributions (see Fig.~\ref{fig:ccqe}) in $\nu_\mu$ events with zero pions in the ME dataset indicate that the region selected by the neutrino-electron scattering analysis (very low lepton $P_t$ and low recoil) are more modestly overpredicted, while the region occupying the sidebands (moderate lepton $P_t$ and low recoil) are well predicted. The remaining discrepancy seen by Ref.~\cite{Ruterbories:2018gub} populates regions of higher hadronic recoil energy (correlated with large vertex energies), as seen in Fig. 38 of Ref.~\cite{Ruterbories:2018gub}, which is not relevant to this result.
Because the overprediction at low $P_t$ and low recoil is comparable to the size of the uncertainty from the RPA correction, no additional uncertainty is assessed.  
In addition to the above described uncertainty on the $Q^2$ shape, a 10\% uncertainty on the normalization of the quasi-elastic background is also assumed.  

A previous \minerva~measurement of $\nu_e$ quasi-elastic scattering~\cite{Wolcott1} indicates no significant disagreement between $\nu_e$ and $\nu_\mu$ predictions, so no additional uncertainty is assigned to account for the extrapolation of the \minerva~$\nu_\mu$ quasi-elastic measurement to the predicted $\nu_e$ spectrum.   The total contribution to the neutrino-electron scattering rate uncertainty from elastic and quasi-elastic sources is 1.0\%.  This includes a negligible contribution from the model of neutrino-electron scattering itself, assessed using an alternate calculation of radiative corrections to the GENIE tree-level calculations~\cite{Sarantakos:1982bp,Bahcall:1995mm}.  

A large source of interaction-model uncertainty arises from predictions of neutral-current $\pi^0$ events, such as diffractive and coherent pion production.  Most of these events are vetoed by the cut on $\mathrm{dE/dx}_{\langle4\rangle}$, but some survive.  The $\mathrm{dE/dx}_{\langle4\rangle}$ sideband described in Sec.~\ref{sec:background} contains a significant underprediction of events, which is attributed to neutral-current coherent events in the background constraint procedure.  A separate fit is also performed that ascribes that overprediction to neutral-current diffractive $\pi^0$ production and takes the difference between the two as a systematic. GENIE uncertainties associated with neutral-current pion production, such as the axial vector mass for resonance production, also lead to uncertainty in this analysis, with all pion production model uncertainties yielding a 0.5\% uncertainty on the neutrino-electron scattering rate.  Uncertainty in models of pion and proton final-state interactions in the primary nucleus each contribute an additional 0.5\% uncertainty.  Other sources of model uncertainty, primarily arising from uncertainties propagated from the background fits, contribute 0.7\% uncertainty to the neutrino-electron scattering rate.   

\subsection{Detector Mass Uncertainty}

Finally, the mass assay of the \minerva~ tracker carries a 1.4\% uncertainty.  While this is technically an uncertainty on the simulated prediction of the number of neutrino-electron scatters in \minerva, rather than on the \minerva~ measurement itself, it is included in the total uncertainty budget of the data spectrum to facilitate the flux constraint described in the following section.  

\begin{table}[ht]
\centering
\begin{tabular}{ c | c }
\hline \hline
Source & Uncertainty (\%) \\ \hline
Beam &  0.21 \\
Electron Reconstruction & 0.57 \\
Interaction Model &  1.68 \\
Detector Mass &  1.40 \\ \hline
Total Systematic &  2.27 \\ \hline
Statistical &  4.17 \\ \hline
Total &  4.75 \\ 
\hline \hline
\end{tabular}
\caption{Uncertainties on total number of neutrino-electron scatters in \minerva, after background subtraction and efficiency correction.  }
\label{tab:error_summary}
\end{table}

\begin{table}
\centering 
\begin{tabular}{ c| c c c c c c}
\hline
\hline
$E_e$ range (GeV)  & 0.8-2 & 2-3 &3-5 &5-7 & 7-9 & 9-$\infty$  \\ 
\hline
 Events  & 329.68&200.88&310.05&167.62&78.77&101.47\\ 
   & 30.63&19.40&23.64&18.01&12.76&19.62\\ 
\hline
 0.8-2  & 938.15&31.55&27.73&16.00&8.22&17.51\\ 
 2-3  & 31.55&376.23&16.91&9.71&1.46&5.60\\ 
 3-5  & 27.73&16.91&558.98&16.84&9.02&17.25\\ 
 5-7  & 16.00&9.71&16.84&324.21&9.62&16.88\\ 
 7-9  & 8.22&1.46&9.02&9.62&162.75&18.19\\ 
 9-$\infty$  & 17.51&5.60&17.25&16.88&18.19&384.95\\ 
\hline
\end{tabular}
\caption{Number of neutrino-electron scattering events in each bin of neutrino energy after background subtraction and efficiency correction, with their uncertainties and covariance.  }
\label{tab:results}
\end{table}

\section{Flux Constraint}
\label{sec:flux}

\begin{figure}[ht]
\centering
  \includegraphics[width=0.45\textwidth]{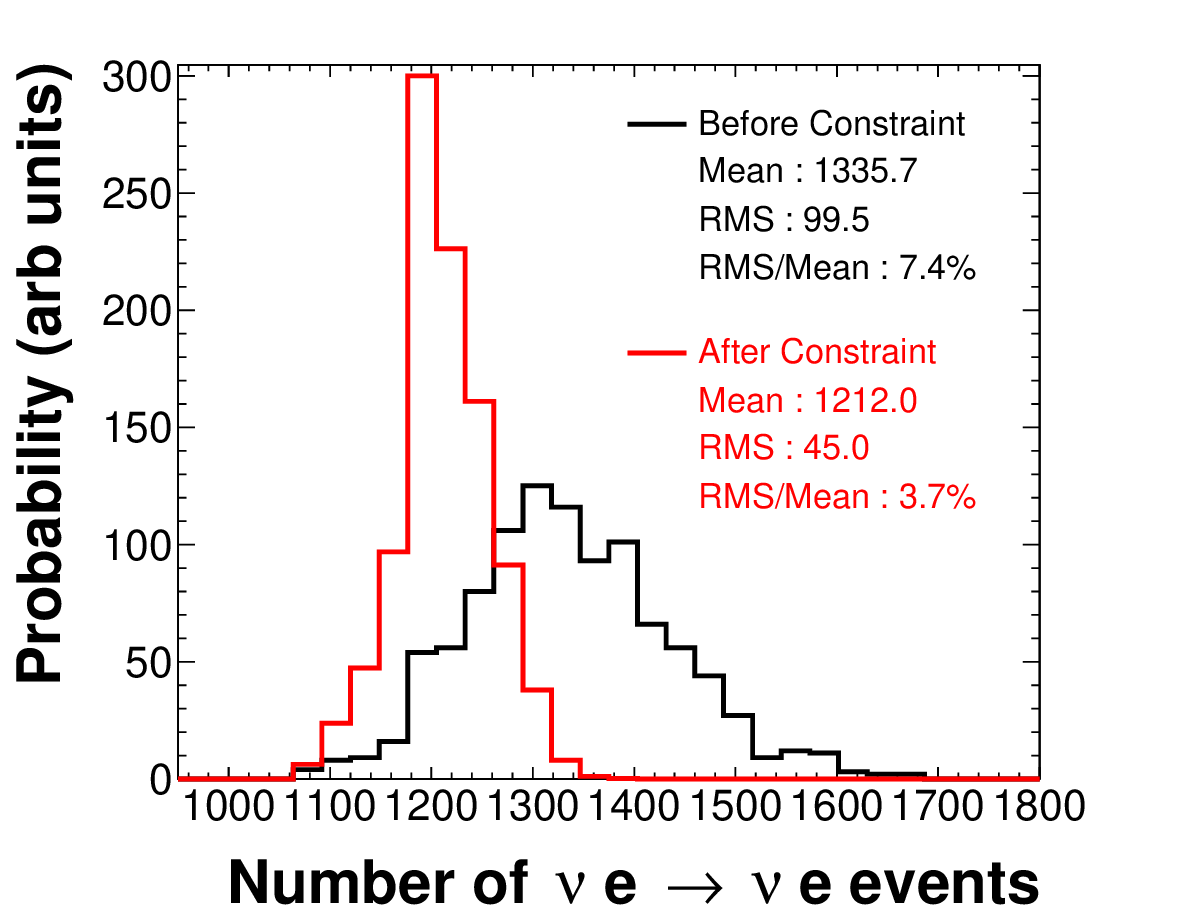} 
\caption{Probability distributions of the number of neutrino-electron scatters, before and after constraining the {\it a priori} flux model using the neutrino-electron scattering data.}  
\label{fig:prob_dist_1}
\end{figure}

The neutrino-electron scattering measurement described here is combined with the {\it a priori} ME NuMI flux prediction at \minerva~in a manner similar to that used for the previous \minerva~ measurement in the NuMI LE beam~\cite{Park:2015eqa}.  This technique relies on Bayes' theorem~\cite{Tanabashi:2018oca}, which holds that the probability of a model given a measurement is proportional to the product of the probability of the measurement given the model and the prior probability of the model.  Thus, a probability distribution of neutrino flux that takes into account both the {\it a priori} flux uncertainties and the neutrino-electron scattering measurement can be obtained by weighting the {\it a priori} probability distribution by the likelihood of the neutrino-electron scattering data given models in that distribution.  

\begin{figure}[hb]
\centering
  \includegraphics[width=0.45\textwidth]{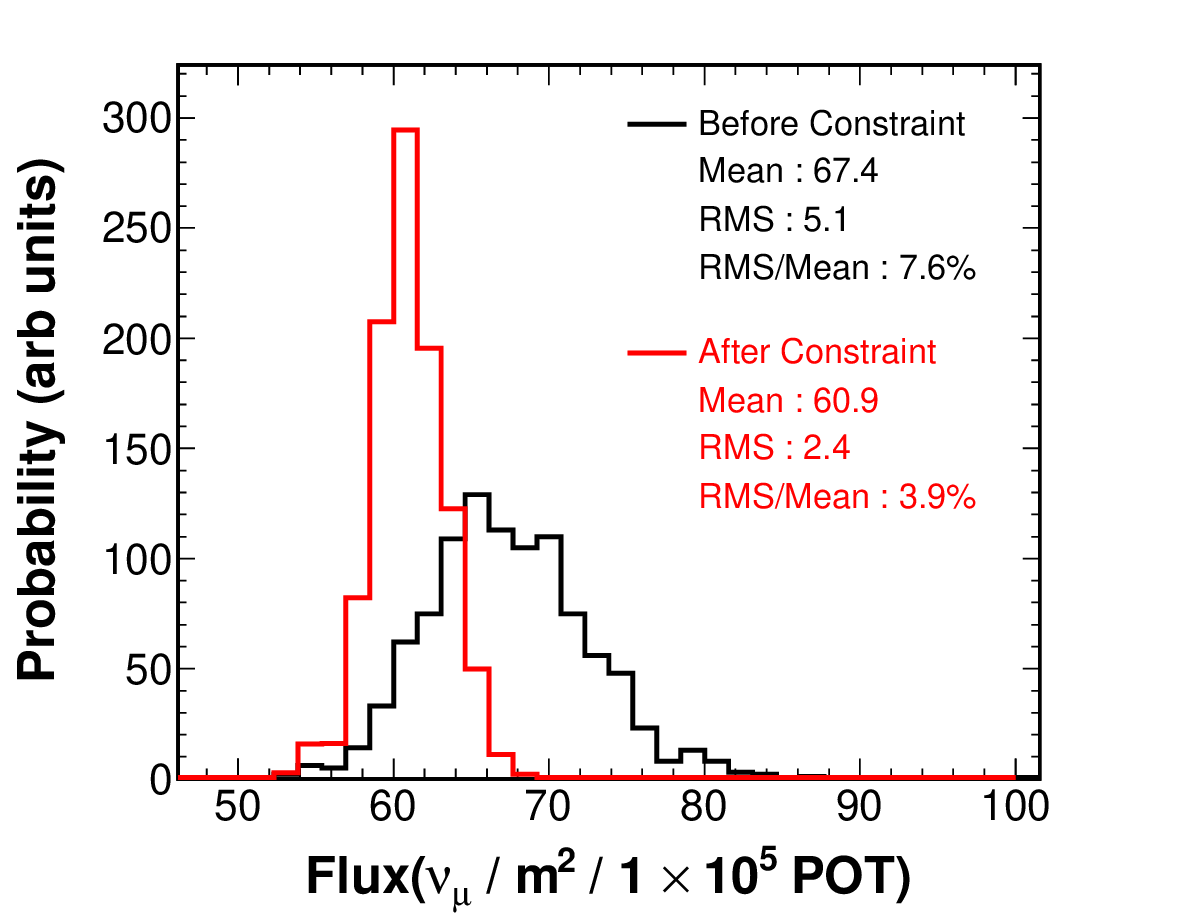}
\caption{Probability distributions of the $\nu_\mu$ flux between 2 and 20 GeV, before and after constraining the {\it a priori} flux model using the using the neutrino-electron scattering data.  }
\label{fig:prob_dist_2}
\end{figure}

In practice, an ensemble of neutrino flux predictions is created with the underlying uncertain neutrino flux parameters (arising from hadron production models and beam alignment) varied within their uncertainties taking into account correlations between parameters.  Each simulation produces a prediction of the number of neutrino-electron scatters in \minerva~ (as well as a variety of other quantities).  To construct the probability distribution for some simulated observable constrained by neutrino-electron scattering data, each simulation is weighted using a $\chi^2$-based likelihood:
\begin{equation}
\label{eq:fluxweight}
W = \frac{1}{(2\pi)^{K/2}}\frac{1}{|\Sigma_{\mathbf N}|^{1/2}}e^{-\frac{1}{2}\left({\mathbf N}-{\mathbf M}\right)^T\Sigma_{\mathbf N}^{-1}\left({\mathbf N}-{\mathbf M}\right)}
\end{equation}
~\cite{Lyons86statisticsfor}, where $K$ is the number of bins in the measurement, $\mathbf N$ is the vector of the bin contents of the spectrum measured in data, ($\mathbf M$) is a vector holding the contents of the predicted spectrum in the simulation in question, and $\Sigma_N$ is the total data covariance matrix describing all uncertainties on ${\mathbf N}$.  

\begin{figure}[ht]
\centering
  \includegraphics[width=0.45\textwidth]{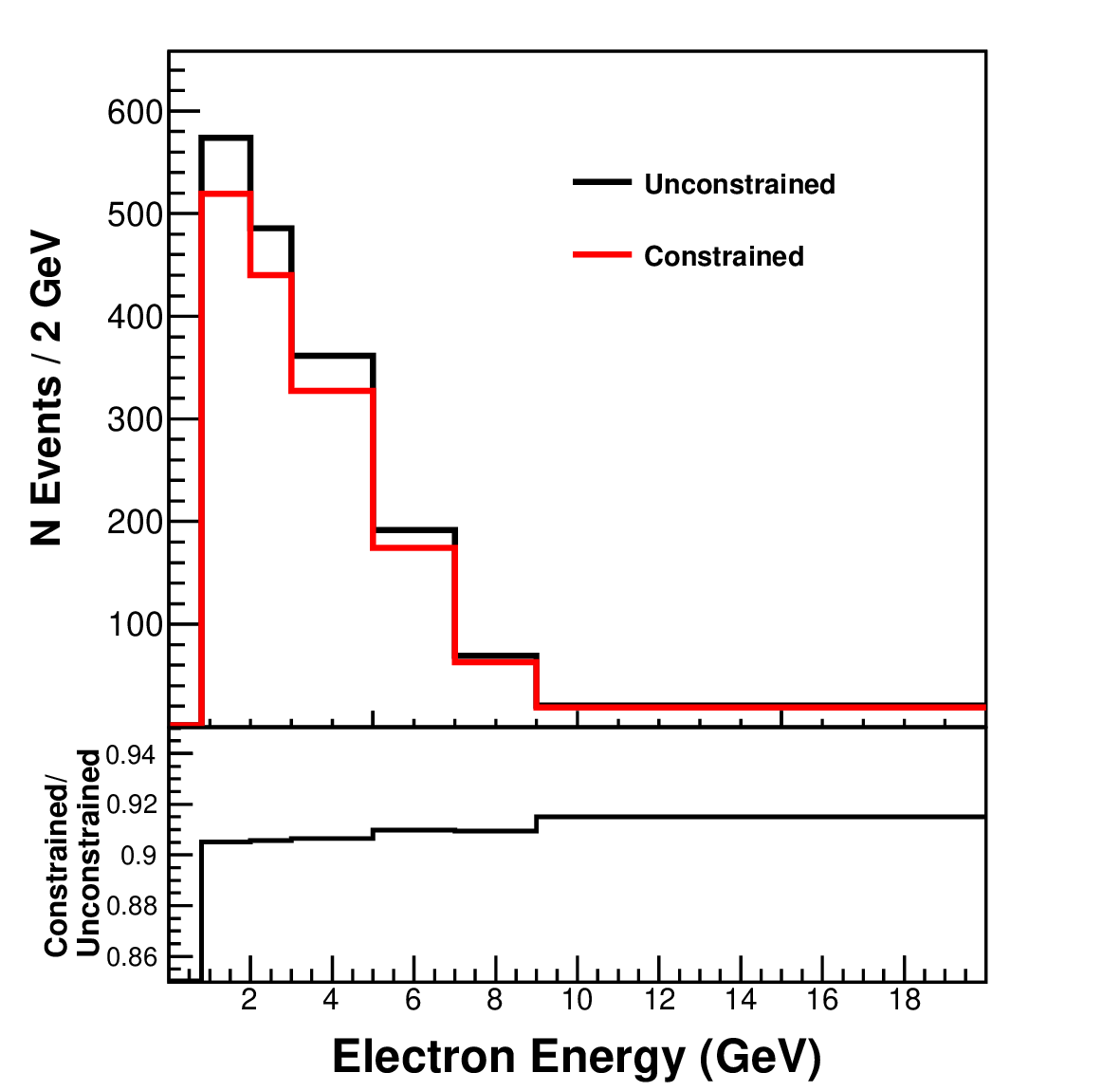} 
\caption{Predicted neutrino-electron scattering electron spectrum  flux, before and after constraining the {\it a priori} flux model using the neutrino-electron scattering data.   }
\label{fig:fluxconstraint_espectrum}
\end{figure}

\begin{figure}[ht]
\centering
  \includegraphics[width=0.45\textwidth]{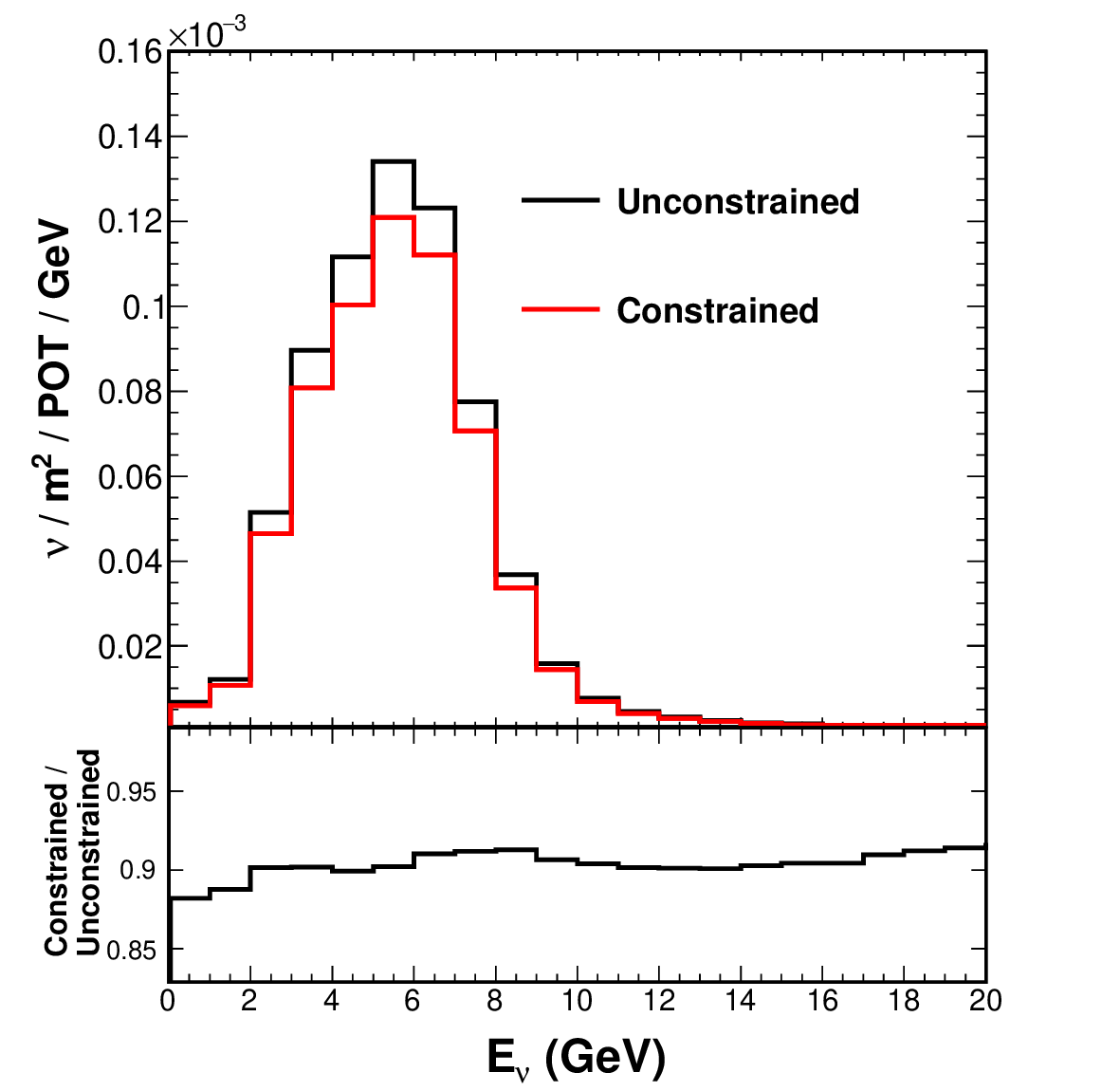} 
\caption{Predicted $\nu_\mu$ flux in bins of neutrino energy, before and after constraining the {\it a priori} flux model using the neutrino-electron scattering data.   }
\label{fig:fluxconstraint1}
\end{figure}

\begin{figure}[ht]
\centering
  \includegraphics[width=0.45\textwidth]{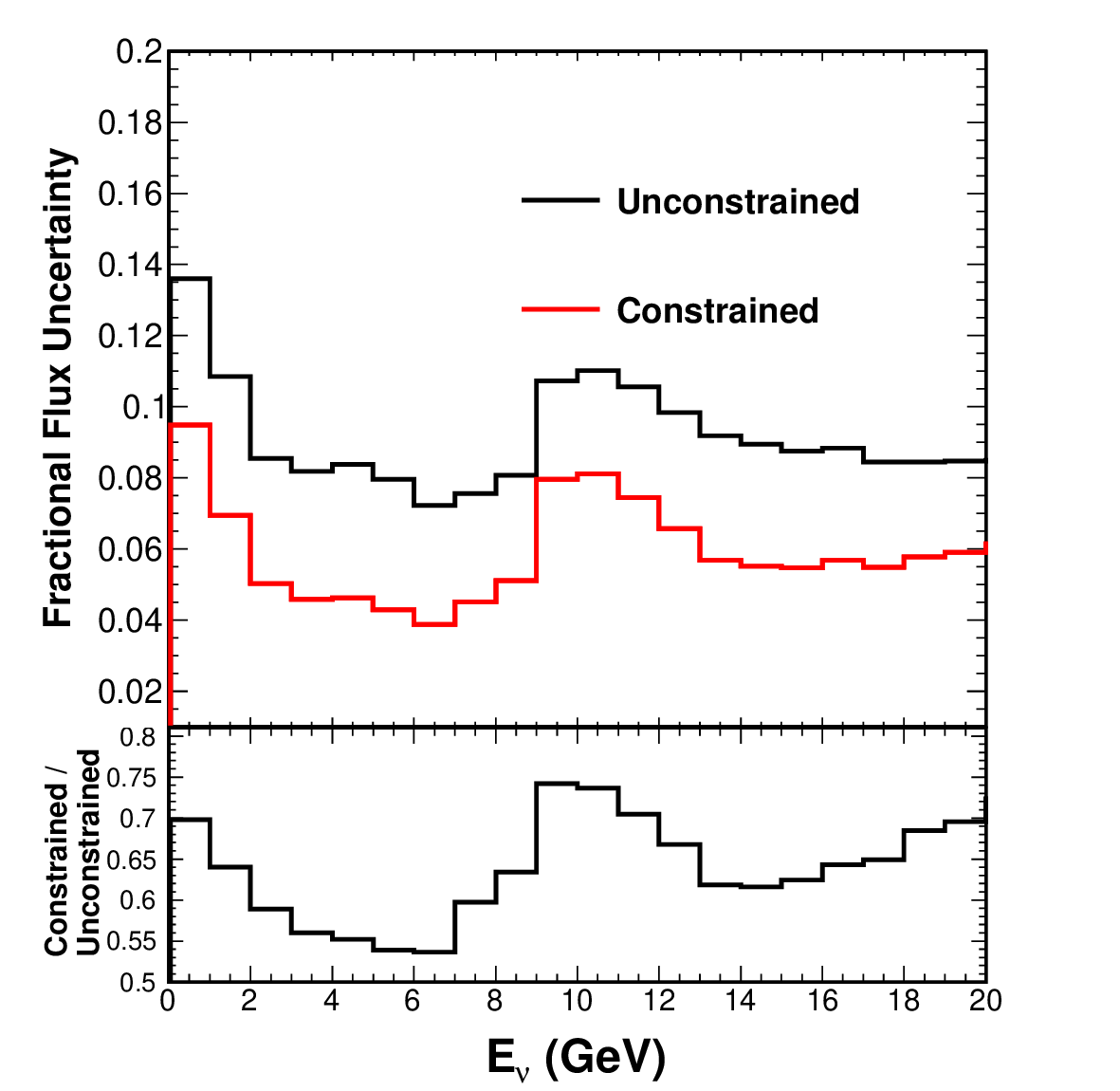}
\caption{Fractional uncertainties on the predicted $\nu_\mu$ flux in bins of neutrino energy, before and after constraining the {\it a priori} flux model using the neutrino-electron scattering data.}
\label{fig:fluxconstraint2}
\end{figure}

The probability distribution (PDF) of the simulated number of neutrino-electron scatters before and after constraint (i.e. applying the weights given in Eq.~\ref{eq:fluxweight}) is shown in Fig.~\ref{fig:prob_dist_1}.  Similar distributions for the predicted number of $\nu_\mu$ between 2 and 20 GeV are shown in Fig.~\ref{fig:prob_dist_2}.  The predicted neutrino flux between 2 and 20 GeV after constraint (taken from the mean of the weighted PDF) is shifted down by 9.6\%, while the uncertainty on the prediction (taken from the RMS of the weighted PDF) is lowered by 53\%.

Similar PDFs can be constructed for the individual bins of any distribution predicted by MINERvA's simulation.  For example, the predicted electron spectrum in neutrino-electron scattering is shown in Figure~\ref{fig:fluxconstraint_espectrum}, and the predicted flux and uncertainty in each bin before and after constraint are shown in Figs.~\ref{fig:fluxconstraint1} and~\ref{fig:fluxconstraint2}, respectively.  Both the electron energy spectrum and the neutrino flux are shifted to slightly higher energy by the constraint.  

\begin{figure}[]
\centering
  \includegraphics[width=0.45\textwidth]{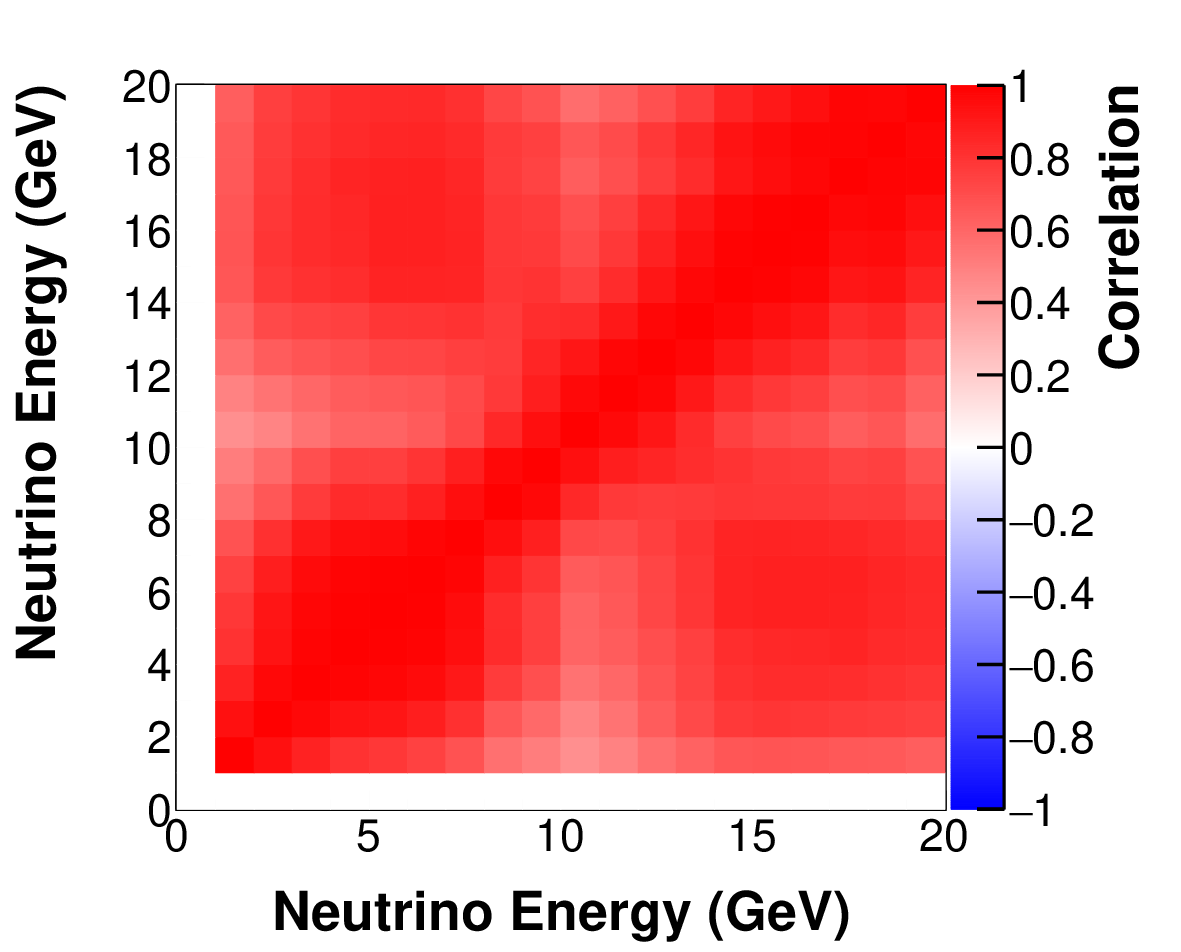}
  \includegraphics[width=0.45\textwidth]{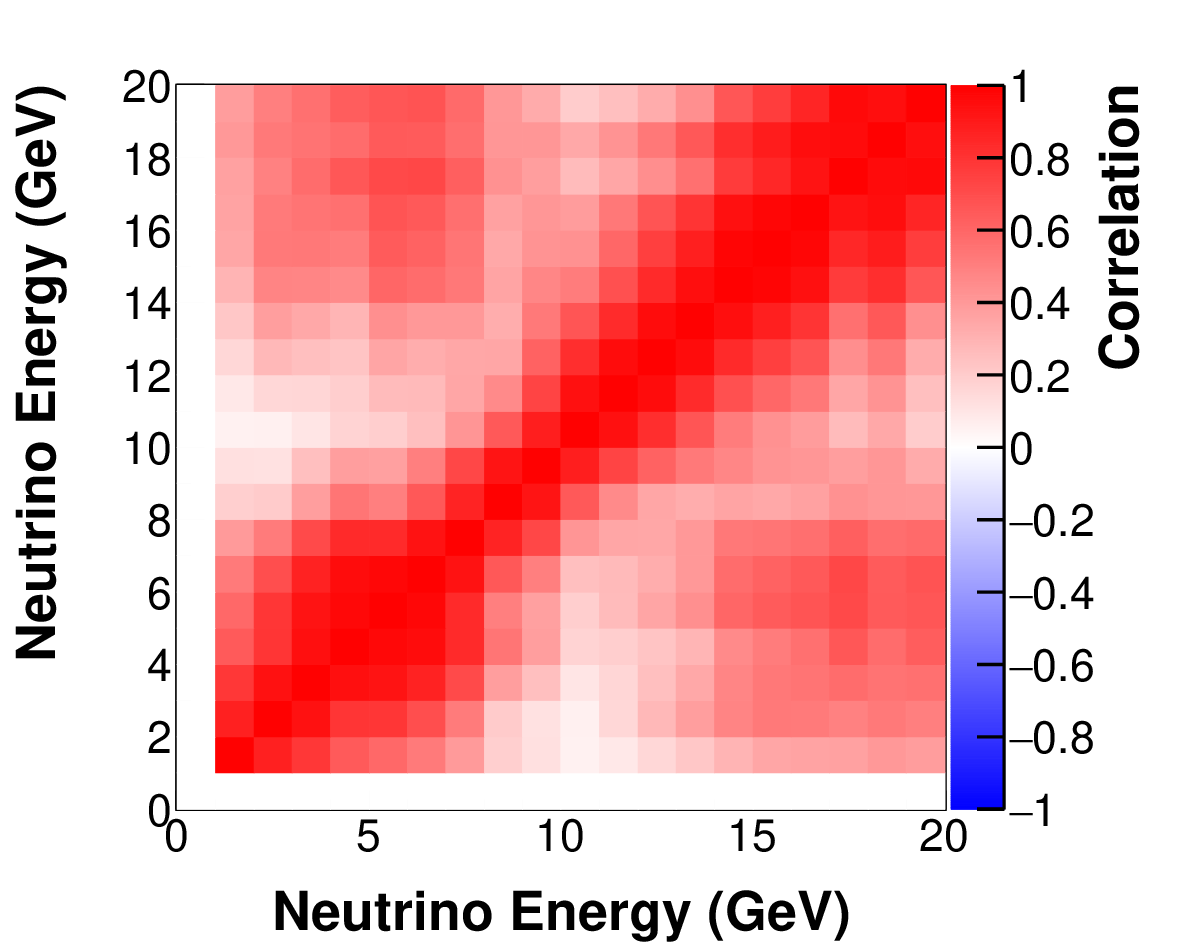}
\caption{Correlation of unconstrained (above) and constrained (below) flux uncertainties.  The uncertainies themselves are shown in in Fig.~\ref{fig:uncertainty}. }
\label{fig:correlation}
\end{figure}

Correlations of uncertainties on the predicted neutrino flux in different energy bins are shown in Fig.~\ref{fig:correlation} for both the unconstrained and constrained cases.  Prior to constraint, the neutrino flux uncertainties are highly correlated across energy bins, due to strong assumed correlations in the model used to predict hadron production off of the target, which dominates the flux uncertainty in most energy regions.  The exception is the 8-12 GeV region, where uncertainties associated with alignment and focusing of the beam are large and uncorrelated with the hadron production uncertainties.  After the constraint, the correlation across neutrino energy bins is weaker.  In particular, the 8-12 GeV region and high energy tail regions are less correlated with the peak region. 

This is the second use of neutrino-electron scattering as a flux constraint by the MINERvA experiment.  The first~\cite{Park:2015eqa} used data from and constrained the flux prediction of the NuMI LE NuMI beam configuration. This paper describes a measurement using the ME NuMI beam, in which the proton target has been pulled upstream of the focusing horn, allowing higher energy mesons to be focused by the horns, resulting in a higher energy neutrino beam.  Because the hadrons that produce neutrinos at~\minerva~  in the ME beam come from generally different (high energy) phase space in the two beam configurations, it is not necessarily expected that the neutrino-electron scattering constraint will have the same effect on both flux predictions.   However, the result of the constraint in both cases is to lower the neutrino flux prediction.  In the LE beam, the flux was lowered by 2-4\% depending on energy, which is a much smaller effect than the ~9-11\% reduction in the ME beam.  That is in part due to the much reduced statistics of the LE measurement, but may also be because of differences in the hadron production model’s predictions of LE and ME hadron phase space.

This flux constraint procedure is only valid if the measured data and {\it a priori} Monte Carlo are statistically compatible.  The MINERvA collaboration deems distributions to be statistically compatible if they agree within three standard deviations.  The chi-square statistic comparing the measurement reported here and the {\it a priori} simulation is 5.1 with 6 degrees of freedom, indicating good agreement.    The total number of predicted events is higher in the simulation by 1.7 standard deviations, which has a 2-sided p value of 9\%.  Both of these calculations use the full data covariance matrix and a covariance matrix for the simulation that includes both statistical and {\it a priori} flux uncertainties.

This measurement could be used to constrain other simulations of the ME NuMI flux (e.g. those used by other experiments operating in the NuMI beam).  In that case, the \minerva~ detector can be represented by a hexagonal prism with apothem 88.125 cm
and length 2.53 m, oriented 58 mrad upward from the NuMI beam axis, consisting of $1.99\pm0.03\times10^{30}$
electrons spread uniformly throughout the 5.99 metric ton fiducial mass. The detector is centered 1031.7 m downstream of 
the upstream edge of the first focusing horn in the NuMI
beamline and $-0.240$ m ($-.249$ m) from the beam
horizontal (vertical) axis.

\section{Conclusion}
\label{sec:conclusion}
The article reports the number of neutrino interactions with electrons in the \minerva~scintillator tracker using a data sample corresponding to $1.16\times10^{21}$ protons on target (POT) in the NuMI ME beam.  The total uncertainty on the number of interactions is 4.8\%, which is more than a factor of two lower than the previous measurement of this process in the NuMI beam.  Combined with the \minerva~flux model and uncertainties, this measurement lowers the predicted $\nu_\mu$ flux by 9.6\% and lowers uncertainties at the focusing peak from 7.6\% to 3.9\%.  This improved flux prediction will benefit all \minerva~ cross section measurements that use this data sample.  The path described here can also be followed by future experiments such as DUNE.

\begin{acknowledgments}
This document was prepared by members of the MINERvA Collaboration using the resources of the Fermi National Accelerator Laboratory (Fermilab), a U.S. Department of Energy, Office of Science, HEP User Facility. Fermilab is managed by Fermi Research Alliance, LLC (FRA), acting under Contract No. DE-AC02-07CH11359.
These resources included support for the \minerva~ construction project, and support
for construction also
was granted by the United States National Science Foundation under
Award No. PHY-0619727 and by the University of Rochester. Support for
participating scientists was provided by NSF and DOE (USA); by CAPES
and CNPq (Brazil); by CoNaCyT (Mexico); by Proyecto Basal FB 0821, CONICYT PIA ACT1413, Fondecyt 3170845 and 11130133 (Chile); by CONCYTEC, DGI-PUCP, and IDI/IGI-UNI (Peru); and by the Latin American Center for Physics (CLAF).  We thank the MINOS Collaboration for use of its near detector data. 
We thank the staff of Fermilab for support of the beam line, the detector, and computing infrastructure.
Finally, we are grateful to Oleksandr Tomalak for gently pointing out an error in the radiative correction formulae in the Appendix.

%
%
%
%
%
%
%
%
%
%
%

\end{acknowledgments}

\bibliographystyle{apsrev4-2}
\bibliography{fluxconstraint}

\appendix
\section{Corrections to the GENIE Neutrino-Electron Scattering Model}
\label{app:radcorrections}
At tree level, the neutrino-electron scattering cross section is given by 
\begin{widetext}
\begin{equation}
\frac{d\sigma(\nu e^-\to\nu e^-)}{dy} = \frac{G^2_F s}{\pi}\left[
  C_{LL}^2+C_{LR}^2(1-y)^2-C_{LL}C_{LR}\frac{m y}{E_\nu}\right] ,
\label{eqn:tree-xsec}
\end{equation}
\end{widetext}
where $E_\nu$ is the neutrino energy, $s$ is the Mandelstam
invariant representing the square of the total energy in the center-of-mass frame, $m$ is the electron mass and $y=T_e/E_\nu$, where $T_e$ is the kinetic energy of the final state electron. 
This analysis uses GENIE version 2.12.6 to model the neutrino-electron scattering signal, corrected for updated electroweak couplings, $C_{LL}$ and $C_{LR}$ ~\cite{Erler:2013xha} and one-loop electroweak radiative corrections as calculated in Ref.~\cite{Bardin:1983yb}.  

This is a different calculation  of  $d\sigma(\nu e^-\to\nu e^-)/dy$ than that used in MINERvA's previous flux constraint based on neutrino-electron scattering~\cite{Park:2015eqa}.  In the previous calculation, based on Refs.~\cite{Sarantakos:1982bp,Bahcall:1995mm}, $y$ was defined as $T_e/E_\nu$.  However, for approximately collinear real photon radiation, the MINERvA detector is most likely to measure the sum of the energy of the electrons and real photons.  The calculation of Ref.~\cite{Bardin:1983yb} defines $y\equiv (T_e+E_\gamma)/E_\nu$, and is therefore more representative of what is observed in the detector.  

One deficiency of the calculation of Ref.~\cite{Bardin:1983yb} is that it does not contain the term in the one-loop cross-section proportional to $C_{LL}C_{LR}$.  This deficiency is corrected in a recent calculation~\cite{Tomalak:2019ibg}, and that result is given below.  However, as illustrated in Eqn.~\ref{eqn:tree-xsec} this term also contains an additional power of $m/E_\nu$ compared to the terms proportional to $C_{LL}^2$ and $C_{LR}^2$, and the entire term is therefore negligible at the few-GeV neutrino energies of the \minerva~experiment.

Table~\ref{tab:ewk-couplings} gives the values of the couplings used in GENIE and in this analysis.
\begin{table}[b]
\begin{tabular}{c|ccc}
   & $C_{\text{LL}}^{\nu_ee}$ & $C_{\text{LL}}^{\nu_\mu e}$ & $C_{\text{LR}}^{\nu e}$ \\ \hline
 \text{GENIE 2.6.2} & 0.7277 & -0.2723 & 0.2277 \\
 \text{One loop} & 0.7276 & -0.2730 & 0.2334 \\
\end{tabular}
\caption{Electroweak couplings in GENIE and in a one-loop calculation of $\nu e^-$ elastic scattering.  }
\label{tab:ewk-couplings}
\end{table}
The one-loop cross section, with $y$ defined as above, is
\begin{widetext}
\begin{eqnarray}
\frac{d\sigma(\nu_\ell e^-\to \nu_\ell e^-)}{dy}=&\frac{G_F^2 s}{\pi}&\left[
   \left( C_{LL}^{\nu_\ell e}\right) ^2 (1+\frac{\alpha_{EM}}{\pi}X_1) +\left( C_{LR}^{\nu e}\right) ^2 (1-y)^2 (1+\frac{\alpha_{EM}}{\pi}X_2) \right. \nonumber \\
&&\left. -\frac{C_{LL}^{\nu_\ell e} C_{LR}^{\nu_ e} m y}{E_\nu}(1+\frac{\alpha_{EM}}{\pi}X_3)\right] ,
\end{eqnarray}
\end{widetext}
\noindent   where the $X_i$ correction terms are
\begin{widetext}
\begin{eqnarray}
X_1&=&-\frac{2}{3} \log \left(\frac{2 y E_{\nu }}{m}\right)+\frac{y^2}{24}-\frac{5 y}{12}-\frac{\pi ^2}{6}+\frac{23}{72} \nonumber \\
X_2&=&-\frac{2}{3} \log \left(\frac{2 y E_{\nu }}{m}\right)-\frac{y^2}{18(1-y)^2}-\frac{\pi ^2}{6}-\frac{2y}{9(1-y)^2}+\frac{23}{72(1-y)^2}
\nonumber \\
X_3&=&-\frac{3}{2} \log \left( \frac{2 y E_{\nu}}{m}\right) +\frac{1}{4}+\frac{3}{4y}-\frac{3}{4y^2}-\frac{\pi^2}{6}.
\end{eqnarray}
\end{widetext}

\end{document}